ARTICLE TYPE

# Ensemble-based, large-eddy reconstruction of wind turbine inflow in a near-stationary atmospheric boundary layer through generative artificial intelligence

Alex Rybchuk[1] | Luis A. Martínez-Tossas[1] | Stefano Letizia[1] | Nicholas Hamilton[1] | Andy Scholbrock[1] | Emina Maric[1] | Daniel R. Houck[2] | Thomas G. Herges[2] | Nathaniel B. de Velder[2] | Paula Doubrawa[1]

[1]National Renewable Energy Laboratory, Colorado, USA

[2]Sandia National Laboratories, New Mexico, USA

**Correspondence**
Corresponding author Alex Rybchuk.
Email: alex.rybchuk@nrel.gov

**Abstract**

To validate the second-by-second dynamics of turbines in field experiments, it is necessary to accurately reconstruct the winds going into the turbine. Current time-resolved inflow reconstruction techniques estimate wind behavior in unobserved regions using relatively simple spectral-based models of the atmosphere. Here, we develop a technique for time-resolved inflow reconstruction that is rooted in a large-eddy simulation model of the atmosphere. Our "large-eddy reconstruction" technique blends observations and atmospheric model information through a diffusion model machine learning algorithm, allowing us to generate probabilistic ensembles of reconstructions for a single 10-min observational period. Our generated inflows can be used directly by aeroelastic codes or as inflow boundary conditions in a large-eddy simulation. We verify the second-by-second reconstruction capability of our technique in three synthetic field campaigns, finding positive Pearson correlation coefficient values ($0.20 > r > 0.85$) between ground-truth and reconstructed streamwise velocity, as well as smaller positive correlation coefficient values for unobserved fields (spanwise velocity, vertical velocity, and temperature). We validate our technique in three real-world case studies by driving large-eddy simulations with reconstructed inflows and comparing to independent inflow measurements. The reconstructions are visually similar to measurements, follow desired power spectra properties, and track second-by-second behavior ($0.25 > r > 0.75$).

## 1 | INTRODUCTION

Even[‡] though gigawatts of utility-scale wind turbines have been deployed to date, it remains difficult to precisely assess the accuracy of the tools used to design those turbines when they operate within the atmospheric boundary layer (ABL). While many factors complicate the model validation and uncertainty quantification process for utility-scale turbines, here we are particularly interested in the challenges associated with the turbulent inflow, as inaccurate inflow reconstruction hampers[1] model validation for complicated turbine dynamics that can occur for large turbines[2] (e.g., unsteady aerodynamics, vortex-induced vibrations). In this study, we are specifically interested in an accurate and probabilistic reconstruction of the ABL three rotor diameters ($D$) upwind of the turbine over a single 10-min window. While utility-scale turbine validation studies often compare against time-averaged quantities of interest[1,3,4], in this paper, we are also motivated by time-resolved turbine validation (also known as one-to-one turbine validation) in which one validates against time-varying quantities of interest to

---

**Abbreviations:** ABL, atmospheric boundary layer; LER, large-eddy reconstruction; LES, large-eddy simulation; OSSE, observing system simulation experiment; RAAW, Rotor Aerodynamics Aeroelastics and Wake.

[‡] This Work has been submitted to Wind Energy. Copyright in this Work may be transferred without further notice

specifically examine dynamical behavior. As such, we seek to produce inflows that track observations on a second-by-second basis, as opposed to those that strictly track time-averaged quantities[5].

To reconstruct second-by-second inflow upwind of a turbine, three components are needed: wind observations, a model of atmospheric winds, and an analysis technique to reconcile the two. Wind observations often come from a meteorological mast[6], though increasingly, nacelle-mounted lidar systems[7,8,9] and hub-mounted lidar systems[10] are also being used. Observations provide wind information at a small region in space; in isolation, they rarely provide all the inflow information that is needed to run aeroelastic simulations. For example, meteorological masts can provide high-frequency measurements of three velocity components ($u, v, w$), although only at a few fixed points across their vertical span. Conversely, turbine-mounted lidars measure incoming flow across a larger volume, although only in the form of line-of-sight velocity and typically at a coarser temporal resolution. In contrast, atmospheric models that are used for turbine inflows can simulate plausible wind behavior at heights everywhere between the ground and the rotor tip at high temporal resolution, and they typically output all three velocity components that are used by turbine aeroelastic codes. However, when they are not coupled to observations or mesoscale models, atmospheric turbine inflow models simulate idealized atmospheric flows that correspond to desired bulk characteristics (e.g., time-averaged hub-height wind speed, turbulence intensity (TI) over a 10-min window) as opposed to an observed, real-world atmosphere. To reconstruct second-by-second, real-world inflow over the full rotor disk, analysis techniques combine information from both observations and models. In the wind energy community, these blending techniques have been based off methods such as kriging[11,12] and linear interpolation within the spectral domain[13]. This process of reconstructing turbine inflows by combining observations with atmospheric modeling is often referred to as "constrained turbulence generation" within the wind energy community.

To date, one-to-one turbine validation studies[6,7,8,14,15] have used practical, but physically limited, spectral-based models of turbulence to generate observation-informed, second-by-second inflows. Turbine inflows are predominantly modeled by using one of two spectral-based models of turbulence: Kaimal-based methods[16] and Mann-based methods[17]. Spectral-based turbulence models come with a number of benefits. For example, they are relatively computationally inexpensive, they are trusted enough to be incorporated into design standards[18], and—of particular interest here—they have demonstrated that they can be used to reconstruct turbine inflow with second-by-second measurements from point sensors and scanning lidars[13,19,20].

While spectral-based turbulence models are powerful and have demonstrated success in the wind sector, they come with several limitations for use in one-to-one turbine model efforts. First, while these inflow models have already been used for one-to-one turbine validation efforts, the constrained turbulence models themselves have never been validated in their ability to reconstruct second-by-second inflows against atmospheric measurements. While inflow generation models have been validated for time-averaged statistics[7], to our knowledge there are no published studies in which a constrained inflow is created by using atmospheric observations at one region in space (e.g., hub height) and then validated against held out data in a second-by-second basis at a different location (e.g., at the top of the rotor). Reconstructed atmospheric winds are a key input into the one-to-one turbine validation model process, and as such, it is important to characterize the accuracy and confidence associated with these reconstructions. Second, the spectral-based turbulence models use modeling assumptions that limit their physical realism, and these assumptions translate to impacts in modeled turbine loads. For example, Holtslag et al. (2016)[21] discuss the limited coupling between wind shear and turbulence during non-extreme conditions in spectral-based turbulence models; this coupling has implications for fatigue predictions. In a similar vein, spectral-based turbulence models do not simulate the dynamic interactions between buoyancy and velocity fields. As such, these models do not explicitly incorporate atmospheric stability when generating turbulence. This assumption impacts the types of turbulent coherent structures that are generated[22,23], thereby impacting the modeled turbine loads[24,25]. Regarding the physical realism of spectral-based turbulence models, a recent survey paper[26] summarizes that "there are several other phenomena that are missing in current [spectral-based] inflow models that likely have a significant impact on design loads."

In contrast, large-eddy simulation (LES) codes model the atmosphere with higher physical consistency. Within wind energy research, LES is routinely used as "ground truth" against which lower-fidelity models are calibrated[27,28]. LES has been used in utility-scale turbine validation efforts, though only in studies that validate against time-averaged quantities of interest[1,29]. We note that while LES is likely more accurate than spectral-based atmospheric models at modeling the ABL, LES is still an imperfect model approach with documented inaccuracies[30,31], and it does not consistently validate better than simpler inflow models in wind energy studies[1,29]. Limited validation of LES as well as large computational costs are major barriers to the adoption of this atmospheric model by the wind industry[32].

Over the years, LES has improved in its ability to simulate ABLs that correspond to real-world conditions. Early wind energy LES studies for finite-length wind farms were relatively idealized. They derived their turbine inflows from precursor LES runs with steady, large-scale forcing and periodic boundary conditions[33,34]. In recent years, wind LES studies have demonstrated the ability to simulate ABLs that better correspond to real-world ABLs by including unsteady large-scale information into inflow-outflow boundary conditions. To date, the time-varying forcing is driven by data that is relatively coarse for LES spatiotemporal scales. Data comes from either mesoscale models[35,36] (grid size $\mathcal{O}(1\text{ km})$) or vertical columns of observations available at 10-min intervals[37,38]. In contrast, LES of the ABL is conducted at spatial resolutions of $\mathcal{O}(10\text{ m})$ and temporal resolutions of $\mathcal{O}(1\text{ s})$. Outside of the wind energy community, we note that the "LASSO" project[39,40] is also making advances in observation-LES coupling in the ABL, though at coarser spatiotemporal timescales than we are interested in here.

While LES codes have advanced over the years, they still have not been used as part of one-to-one turbine validation studies. Fundamentally, this issue stems from the challenge of reconciling turbulent LES fields ($\mathcal{O}(10\text{ m}, 1\text{ s})$) with second-by-second observations[41]. To be specific, given noisy, high-frequency, spatially limited observations, how does one generate initial conditions and boundary conditions (and possibly optional body forces) for an LES run that could be used to emulate a particular real-world ABL? And how does one do this in a probabilistic manner to account for the highly chaotic nature of the ABL? One proposed strategy is "data assimilation"[42], a technique deployed routinely by operational weather forecasting centers to fuse information from atmospheric observations and mesoscale simulations or synoptic-scale simulations. In recent years, data assimilation has also been applied to LES studies[43,44], including LES studies of wind energy[45,46]. However, LES-coupled data assimilation has not yet been demonstrated for a real-world fluid flow (e.g., in an experimental facility or in the ABL). While data assimilation produces an observation-informed estimate of the state of the interior of a simulation domain, it does not provide an estimate of boundary conditions. The lack of observation-informed boundary conditions is a major roadblock for LES of an ABL, as the common choice of periodic boundary conditions used for idealized LES simulations does not reflect a real-world ABL. Another strategy to blend observations and simulations of a highly turbulent fluid is "nudging"[47]. The mesoscale-microscale simulations described above use nudging techniques successfully to track flows on relatively coarse temporal scales. However, in a fluid mechanics study of turbulent flow around a cylinder, Zauner et al. (2022)[48] found that nudging a flow with a coarse observation network (as would be found in an ABL field campaign) produces unphysical artifacts.

In recent years, machine learning algorithms have shown promise in their ability to reconstruct the state of highly turbulent fluids from limited measurements, thereby potentially addressing some barriers for LES-based inflow reconstruction. Fluid mechanics studies have demonstrated machine learning-based flow reconstruction in both computational environments[49,50] and in experimental fluid mechanics facilities[51,52]. The weather community is also developing new techniques[53,54] that leverage machine learning to reconcile observations and atmospheric simulations on the mesoscale and synoptic scales. In earlier work[55,56], we studied the ability of one particular class of machine learning algorithm—a diffusion model[57,58]—to reconstruct ensembles of three-dimensional turbulent states from a set of noise-free, spatially limited measurements in a computational analogue of a field campaign. We found that the diffusion models were able to generate plausible realizations of instantaneous atmospheric turbulence, even with relatively sparse measurements. The ensemble of realizations for each individual observation showed meaningful diversity, thereby providing a measure of uncertainty associated with the reconstruction. Crucially, we demonstrated that these reconstructions successfully interfaced with an LES code by treating them as initial conditions and conducting hundreds of numerically stable LES runs. While these studies demonstrated the potential of using diffusion models for real-world turbine inflow reconstruction, they had three fundamental limitations. First, they omitted all time history knowledge from observations, reconstructing an instantaneous box of fluid flow instead of time-varying information. Second, they used idealized, noise-free measurements. And third, they were tested in a "best-case scenario" synthetic LES environment instead of in the real world.

In this paper, we develop, verify, and validate an approach to generate turbine inflow that is grounded in an LES model of the atmosphere and conditioned on second-by-second, real-world measurements. We generate ensembles of inflows by leveraging diffusion models. We verify the accuracy of our inflow generation technique by running three synthetic field campaigns in an LES run, enabling us to identify strengths and weaknesses of the algorithm in a best-case scenario. We then validate our technique in the real world by using measurements from three different time windows of the Rotor Aerodynamics Aeroelastics and Wake (RAAW) field campaign. While our new inflows enable us to conduct one-to-one validation of turbine models with LES, we limit our scope to the turbine inflow and omit validation of the turbine model. We stress that our reconstruction methodology also comes with limitations (detailed later in this manuscript), though an initial one-to-one turbine validation effort suggests that LES-based inflows have potential to improve upon inflows from kinematic models[59]. In Section 2, we discuss the RAAW field campaign and its observations of the atmosphere. In Section 3, we detail our inflow generation technique. In Section 4, we verify our technique in a synthetic field campaign. In Section 5, we validate against real-world measurements and also conduct an additional synthetic campaign to give context to the real-world results. Then, we summarize our results and limitations in Section 6, and finally conclude in Section 7.

## 2 | THE RAAW FIELD CAMPAIGN

The RAAW campaign collected high-resolution and high-coverage atmospheric and turbine measurements in order to validate computational models of the aeroelastic, aerodynamic, and wake behavior of modern land-based wind turbines with flexible blades. The experiment took place in Lubbock, Texas, United States, between May and October 2023. The subject of the experiment was a 2.8-MW wind turbine manufactured and owned by GE Vernova, with a rotor diameter of $D = 127$ m and a hub height of 120 m. The absence of nearby wind turbines and the simple terrain simplifies the task of model validation, allowing us to focus on the detailed dynamics of the inflow and structural response to a single operating turbine without the added complexities of wind farm interactions and topography.

### 2.1 | Instrumentation

The campaign deployed several measurement systems that we use in this study. We primarily reconstruct inflow with the use of a commercial nacelle-mounted scanning lidar (Halo Streamline XR) that faces upwind. Throughout this text, we use the nomenclature "lidar" or "inflow lidar" to specifically refer to this lidar that is used for inflow reconstruction, whereas the SpinnerLidar (discussed more below) is always referred to by its full name. The inflow lidar conducts horizontal plan position indicator (PPI) scans between azimuth values of -15° to 15° at a resolution of 2°, sweeping parallel to the ground at hub height and measuring line-of-sight velocities. The scan was optimally designed to capture the most energetic spatiotemporal modes expected in the flow[60]. The lidar scans at each azimuth for 1 s, and takes 2 s to reset to its initial azimuth after each sweep. As such, one two-dimensional scan takes 18 s to complete. The range gate length of the lidar is 12 m. During high thrust operation, the turbine tower bending induces a positive tilt of ∼1° on the lidar scanning plane, which translates into a vertical displacement of less than 6 m at $3D$, and we neglect this effect in our inflow reconstruction process. As the inflow lidar is mounted on the nacelle and stares upwind, turbine blades sweep by during measurements, obstructing between a third and a quarter of measurements from this instrument.

While the inflow lidar provides the second-by-second observations that we explicitly use for the inflow reconstruction, we use information from two additional instruments to implicitly condition our inflow planes. A meteorological mast sits ∼390 m (∼$3D$) away from the turbine, along the direction 185° relative to north. The mast is instrumented with cup anemometers near the bottom of the rotor (∼56 m), at hub height (120 m), and near the top of the rotor (∼183 m). We use cup anemometer measurements to calculate time-averaged vertical profiles of wind speed and turbulence intensity. We also use a profiling radiometer at the site to estimate the height of the capping inversion.

We validate against measurements from an independent (i.e. not used in the reconstruction) inflow sensor. The hub-mounted SpinnerLidar[61,62] measures line-of-sight winds following a rosette pattern in the SpinnerLidar frame of reference ∼$1D$ upstream with a diameter approximately matching the rotor at 127 m and a maximum half-angle of 30°. The SpinnerLidar conducts high-frequency scans (approximately 977 points every 2 s). Importantly, the SpinnerLidar provides inflow measurements above and below hub-height that the inflow lidar has not seen. As such, we validate our reconstructions against these winds to test their accuracy in regions where the reconstructions did not have observational data.

### 2.2 | Intensive observation periods

We select three 704-s windows during which our key instruments reported high-quality data. The atmospheric dynamics are relatively stationary and consistent across all three periods (Table 1). These periods start at 15:00:25 UTC, 15:30:25 UTC, and 16:00:25 UTC on July 24, 2023, and we refer to these periods as the 1500, 1530, and 1600 Periods, respectively. The measurements were collected during daytime on a clear-sky day (local time is UTC − 5 hours). We select the window lengths to be slightly longer than 10-min periods so that we can discard data during a simulation spin-up transient and then have enough data to validate during the industry-standard, 10-min window. Between 15:00:00 UTC and 16:15:00 UTC, the average hub-height wind speed measured at the meteorological mast is 9.03 m s$^{-1}$ and the average surface heat flux is 0.184 K m s$^{-1}$. The vertical shear of wind speed is low for this period, as the time-averaged wind speed is 8.82 m s$^{-1}$ at 56.6 m and 8.96 m s$^{-1}$ at 183.5 m. The average hub-height wind direction is 224° relative to north, which means that the meteorological mast and lidar systems are seeing similar but distinct second-by-second flow. The capping inversion height is approximately 800 m.

**TABLE 1** Meteorological information for each of the real-world case studies as well as the statistics from the longer window that are used for the LES training dataset generation. We calculate time-averaged wind speed, wind direction, and wind shear coefficient as well as the turbulence intensity from the meteorological mast. We calculate the surface heat flux as well as the Obukhov length using measurements from a sonic anemometer and temperature sensor at 2 m.

|  | Mean 120-m Wind Speed [m s$^{-1}$] | 120-m Turbulence Intensity [%] | Mean 120-m Wind Direction [°] | Wind Shear Coefficient ($\alpha$) | Surface Heat Flux [K m s$^{-1}$] | Obukhov Length [m] |
|---|---|---|---|---|---|---|
| **1500 Period** | 9.42 | 10.8 | 255.6 | 0.064 | 0.145 | -103 |
| **1530 Period** | 8.88 | 11.2 | 220.7 | -0.044 | 0.186 | -82 |
| **1600 Period** | 8.46 | 10.7 | 224.3 | 0.038 | 0.188 | -59 |
| **15:00 UTC–16:15 UTC** | 9.03 | 10.4 | 223.8 | 0.013 | 0.184 | -81 |

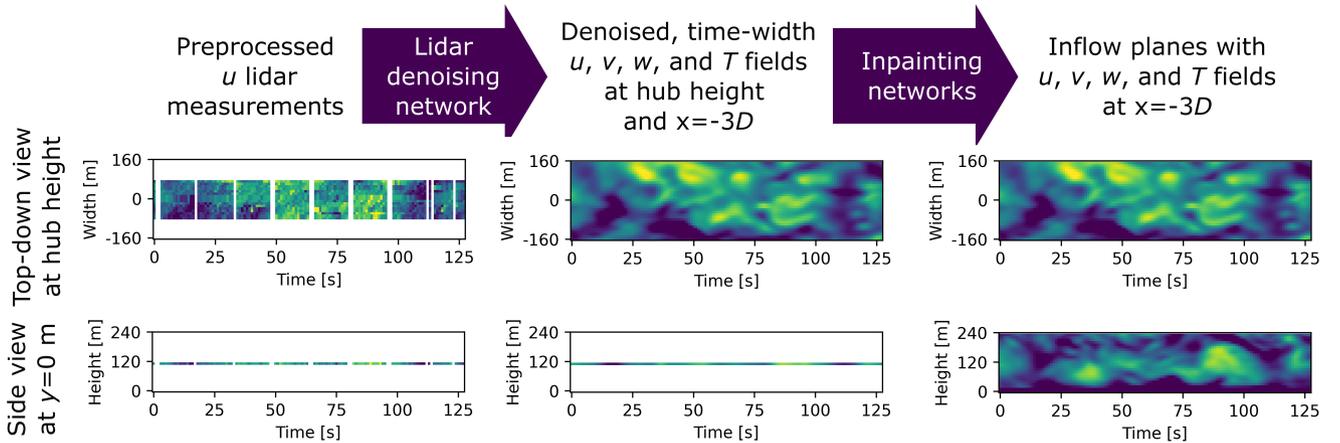

**FIGURE 1** The pipeline used to reconstruct the inflow to the wind turbine. We start from the preprocessed noisy lidar measurements of $u$ at hub height, and we end with a time-varying inflow plane of $u$, $v$, $w$, and $T$ data. See Figure 4 for a more detailed visualization of the inpainting process.

## 3 | INFLOW RECONSTRUCTION METHODOLOGY

### 3.1 | Overview and problem definition

Before detailing the specifics of our inflow reconstruction methodology, we first provide a high-level overview of objectives, inputs, and the process. To distinguish our second-by-second reconstruction problem from the problem of model-observation blending via mesoscale-microscale coupling, we refer to the second-by-second reconstruction problem as "large-eddy reconstruction" (LER). We view the LER problem as a model-observation reconciliation problem, where the goal is to reconstruct large eddies, the kinds of which would be explicitly resolved by an LES code. The reconstructed turbulent flow field could correspond to an instantaneous volume, a time-varying plane, or a time-varying volume. In principle, this problem could be addressed through the use of data assimilation techniques, nudging approaches, or through machine learning algorithms. In this paper, we demonstrate one approach to address the LER problem.

**Problem** In this work, we seek to reconstruct second-by-second inflow 380 m ($\sim 3D$) upwind of a turbine over the course of a 10-min window by blending information from second-by-second observations and an atmospheric model. We select this location to minimize slowdowns from turbine induction effects[63] while maintaining acceptable inflow lidar measurement quality. Our second-by-second observations come from a nacelle-mounted lidar measuring at hub height. Our model of atmospheric winds, which will be used to reconstruct the unobserved inflow information, is based on LES data.

**Inputs** To generate the LES dataset that is used to train the LER networks, we require time-averaged meteorological measurements, namely, vertical profiles of wind speed profiles, vertical profiles of TI, and the height of the capping inversion. To generate the LES dataset, we also require knowledge of the inflow lidar sampling strategy. After the LES dataset has been created and the networks have been trained, we can then generate observation-informed inflows. To generate these inflows, we require line-of-sight measurements from a nacelle-mounted lidar.

**Outputs** The LER pipeline generates an ensemble of $N_{ens} = 30$ inflow fields, each with a width of $L_y = 320$ m, a height of $L_z = 240$ m, and time length $L_{time} = 704$ s at a time step of 1 s and a spatial resolution of 10 m. Therefore, each inflow plane is sized $(N_{time}, N_y, N_z) = (704, 32, 24)$. The reconstruction area covers the entire cross-sectional area of the wind turbine footprint laterally and vertically, plus a buffer region. We reconstruct four correlated variables: three velocity components $(u, v, w)$ and temperature $(T)$. These inflow planes can be used in stand-alone aeroelastic turbine simulations (e.g., OpenFAST) or as boundary condition information in our LES code AMR-Wind[64].

**Process** We convert these inputs into outputs through the use of a preprocessing step as well as two diffusion model architectures (Fig. 1). The first diffusion model reads in preprocessed, gridded $u$ measurements sized $L_{time}$ by $L_{y,in} = 200$ m and outputs a two-dimensional plane of $(u, v, w, T)$ data sized $(L_{time}, L_y)$. In the language of the computer vision literature, this network can be thought of as solving a style transfer problem or more broadly a generic image-to-image translation task, translating from noisy lidar measurements to the underlying LES fields. To use language more akin to the wind energy literature, we refer to this network as the "lidar denoising" network, as it reads in noisy lidar measurements and outputs LES-style fields, though strictly speaking, this network conducts additional operations to noise removal. The second diffusion model architecture reads the two-dimensional output from the first network and vertically extrapolates to the ground and to a height of 240 m to produce three-dimensional inflows sized $(L_{time}, L_y, L_z)$ for all $N_{chan} = 4$ variables of interest. In the computer vision literature, this process would be referred to as either an "inpainting" or "outpainting" task[65,66], during which we fill in regions of missing (unobserved) data in a way that is consistent with our observed data. Correspondingly, we refer to this model as the inpainting model.

## 3.2 | Preprocessing inflow lidar measurements

Instead of directly feeding inflow lidar measurements into the lidar denoising network, we first preprocess them so that the denoising task is simpler. The inflow lidar returns line-of-sight velocity measurements every 12 m along a beam that changes position every 1 s. The measurements from beams during a single left-to-right sweep can be concatenated to obtain a two-dimensional line-of-sight measurement in a sector of a circle at a frequency of 1/18 Hz. We de-project[67] the line-of-sight velocities to estimate streamwise velocity, $u$, perpendicular to the nacelle heading. We follow the temporal upsampling algorithm of Beck and Kühn (2019)[68] to upsample two-dimensional sweeps of $u$ to a temporal resolution of 1 Hz. We adjust their algorithm to account for missing individual beams (due to blade passage) by using a nearest-neighbor interpolation algorithm to fill in missing data in each 1/18-Hz sweep prior to upsampling. After upsampling the sweeps, we use nearest-neighbor interpolation to go from the polar grid (i.e. velocities as a function of azimuth, radius and time) onto a Cartesian grid perpendicular to the rotor (i.e., velocities as a function of streamwise and spanwise spatial coordinates, and time). Only the interpolated measurements 380 m upwind of the nacelle are fed into the machine learning algorithms. Thus, after preprocessing the inflow lidar measurements, we have a rectangular grid of hub-height $u$ measurements sized $(L_{y,in}, L_{time})$.

## 3.3 | Generating a training dataset via a synthetic, LES analogue

After identifying our observational period of interest and the inflow lidar scanning strategy during those periods, we set up a synthetic analogue of the real-world field campaign using LES. This process of running a synthetic field campaign is sometimes referred to as an observing system simulation experiment (OSSE), which is the terminology we adopt here.

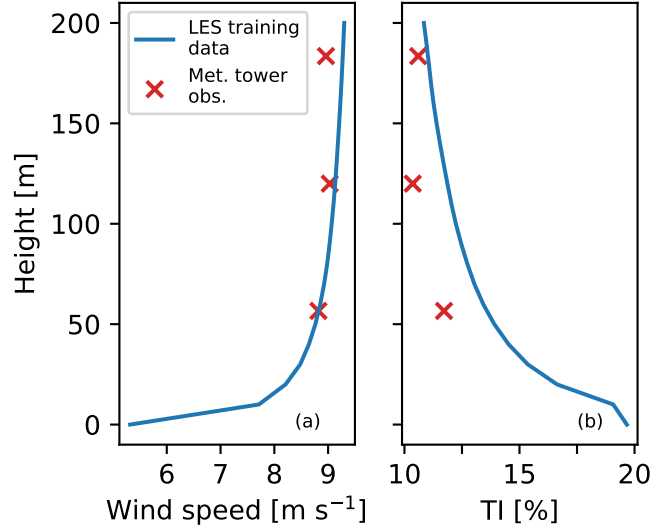

**FIGURE 2** A comparison between (a) the average two-dimensional wind speed and (b) the average TI in the LES training dataset and the meteorological mast measurements between 15:00:00 UTC and 16:15:00 UTC.

We spin up an LES run that approximately matches observed bulk characteristics from the meteorological mast (Fig. 2) as well as the observed capping inversion height. We simulate a flat, horizontally periodic atmosphere in a domain sized (15,360 m, 15,360 m, 1,280 m) with 10-m resolution. We choose this domain size to minimize the development of unphysical periodic structures that occur in small domains[69] while still maintaining a reasonable computational cost. We prescribe the large-scale LES forcing to have a hub-height wind speed of 9.03 m s$^{-1}$ and surface heat flux of 0.184 Km s$^{-1}$. We tune our choice of surface roughness to best match the observed 10-min average wind speeds at the rotor bottom and rotor top, as well as matching the 10-min hub-height turbulence intensity. After sweeping across a range of surface roughness values, we find that a surface roughness of 0.1 m gives us the best match for vertical wind speed shear across the rotor disk and hub-height TI. We take a heuristic tuning approach, as it would be extremely difficult to set up an idealized LES run that exactly matches multiple desired quantities of interest simultaneously to within some desired tolerance.

We allow turbulence to spin up over a period of 3 hours, after which we collect synthetic measurements from a large ensemble of field campaigns over the following hour (Fig. 3). The synthetic field campaigns provide us the data that the diffusion models are trained on. We densely pack the domain with 345 synthetic field campaigns to maximize the size of the machine learning training datasets. In each synthetic field campaign, we collect two pairs of measurements:

1. In the first pair, we collect synthetic lidar measurements that match our real-world scan strategy, and we also collect the true hub-height, time-width plane of measurements 380 m upwind of each lidar. We postprocess[70,71] the synthetic line-of-sight lidar measurements so that they look more like real-world lidar measurements: we add in Gaussian smoothing along range gates, range-dependent noise, and data dropouts due to blade passage. We then upsample and regrid these synthetic lidar measurements in the same manner as the real-world measurements. Thus, we obtain a pair of noisy, upsampled, hub-height $u$ data as well as the true underlying hub-height $u$ fields. This first pair of measurements is used to train the lidar denoising diffusion model.
2. In the second pair of measurements, we collect the true hub-height, time-width plane of measurements 380 m upwind of the lidar as well as the coincident time-width-height box (with spatial dimensions 320 m by 240 m). This pair of measurements is used to train the inpainting diffusion models.

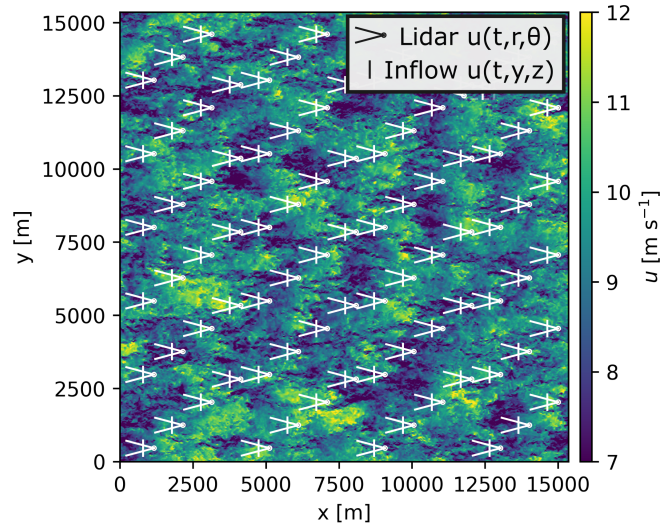

**FIGURE 3** A schematic depicting the synthetic field campaigns, which are repeated throughout the domain to obtain the maximum amount of training data. Each campaign is made up of a lidar system that scans an inflow plane. To make the flow fields visible, only every fourth campaign is plotted.

## 3.4 | The diffusion models: training and sampling

### 3.4.1 | The lidar denoising network

For our lidar denoising process, we train a single diffusion model on our dataset from the ensemble of LES campaigns to translate from one noisy time-width plane of $u$ data to an ensemble of LES-style time-width planes of $(u, v, w, T)$. The details of the model architecture and training are documented in Appendix A. We train on four NVIDIA H100 GPUs. Our architecture requires the input to the network to be the same size as the output of the network. Because the width of our input information $L_{y,in}$ = 200 m is narrower than our desired output width of $L_y$ = 320 m, we reformat our input data to be a masked array sized $(N_{chan}, N_{time}, N_y)$. We center the noisy $u$ information along the $y$-axis and then populate the rest of the masked array with mask fill values (Fig. 1). Once trained, the network produces an ensemble of $N_{ens}$ hub-height reconstructions for a given input.

### 3.4.2 | The inpainting networks

Next, we train a set of three diffusion models to fill in unobserved $(u, v, w, T)$ data around the $(u, v, w, T)$ data from the output of the previous network (see Appendix A for more detail). Unlike with the lidar denoising network, a single inpainting network does not produce the desired output all at once. Instead, we train three networks to read in small chunks of data and to produce slightly larger chunks of data. The first network vertically extrapolates from hub-height data at $z$ = 120 m to data at $z$ = [80, 120] m, the second network vertically expands data from $z$ = [80, 120] m to $z$ = [0, 120] m, and the third network vertically expands data from $z$ = [0, 120] m to $z$ = [0, 240] m. We found that this sequential vertical extrapolation process was necessary. If we attempted to translate directly from $z$ = 120 m inputs to $z$ = [0, 240] m outputs, we found that the network would entirely ignore the input data. Just as with the lidar denoising network, we reformat the input to each network using masked arrays so that the shape of the input data matches the shape of the output data. Similarly, the networks do not process all $N_{time}$ = 704 s of time data at once. Instead, each network processes 256 s of data at a time because we found substantially worse reconstruction skill with longer time windows. We note that each of the three networks is trained to process two different formats of input data: either 256 s of information strictly at the input height, or a combination of this data with 128 s of data at the output height. These two data formats are demonstrated more clearly in the following paragraph.

After training the three networks, we incrementally sample from them (Fig. 4) in order to generate a full box of inflow data. We begin by reading in the first 256 seconds of hub-height data (Fig. 4a); then, by using the smallest inpainting network, we extrapolate downward to 80 m for this period (Fig. 4b). Next, we slide the reconstruction window forward by 128 s. We read in both hub-height data as well as the last 128 s of the previously reconstructed data in order to generate the next 128 s of inflow for this height span (Fig. 4c). This process is repeated between 80 and 120 m until we reconstruct 704 s of inflow. Afterward, we return to the first 256 s (Fig. 4d), and we use reconstructed inflows between 80 m and 120 m to inpaint down to the surface (Fig. 4e) by using the next diffusion model network. Then, we again repeatedly slide the reconstruction window forward 128 s at a time until all 704 s of inflow are reconstructed below hub height. Finally, we repeat this process aloft in order to reconstruct up to 240 m. In principle, this sequential sampling process could produce unphysical artifacts (e.g., seams, discontinuities) between adjacent vertical heights or time windows. However, we do not find any of these artifacts. Our networks here learned the ability to "copy-paste" the input fields directly into the output field (a skill they lacked in our preceding paper[55]), which likely significantly helps mitigate these artifacts.

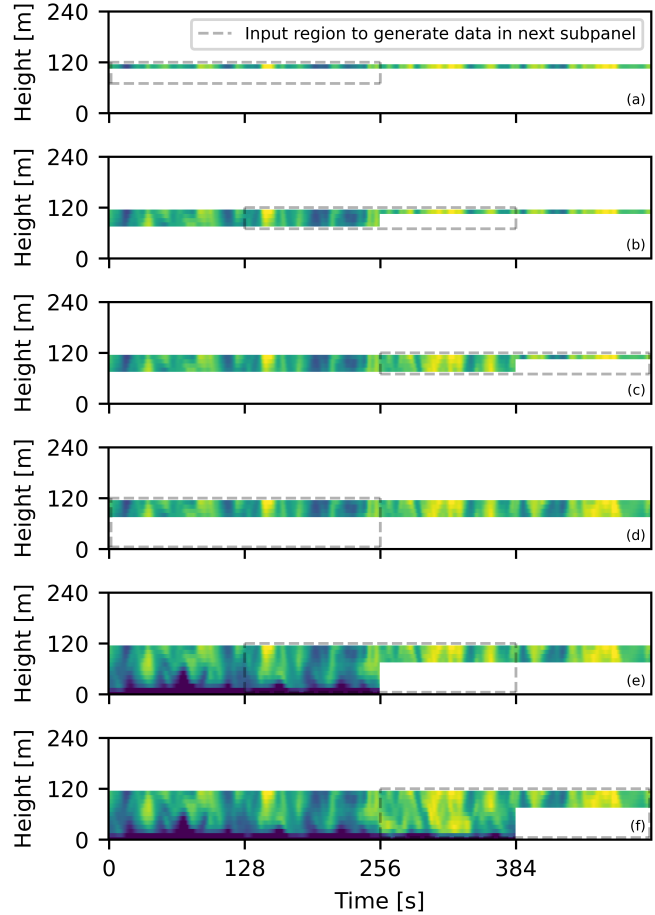

**FIGURE 4** A series of vertical cross sections that demonstrate a portion of the incremental process by which we generate inflows. This incremental inpainting process is repeated above hub height as well, enabling us to generate inflows 704 s long and up to a height of 240 m.

## 4 | INFLOW VERIFICATION AGAINST SYNTHETIC FIELD CAMPAIGNS

### 4.1 | OSSE configuration

While our ultimate goal is to demonstrate the accuracy of our LER approach against real-world measurements, we first assess its performance in an OSSE study that represents our best-case scenario for real-world reconstruction. We spin up a new LES ABL that has the same bulk forcing as the one used to generate training data (Section 3.3), except we initialize the atmosphere with a different perturbation. This change means that our OSSE produces similar turbulent structures that are observed in our training dataset, though the exact timing, location, and shape of those structures is different in the training data generation run. We allow this ABL to spin up for 3.5 h, as this timing places us in the middle of the training dataset generation period. Following a common procedure for running LES of turbines[33], we continue to simulate the atmosphere after spin-up for 15 min using the same periodic boundary conditions that were used during spin-up. During this precursor phase, we collect boundary condition information for the simulation domain.

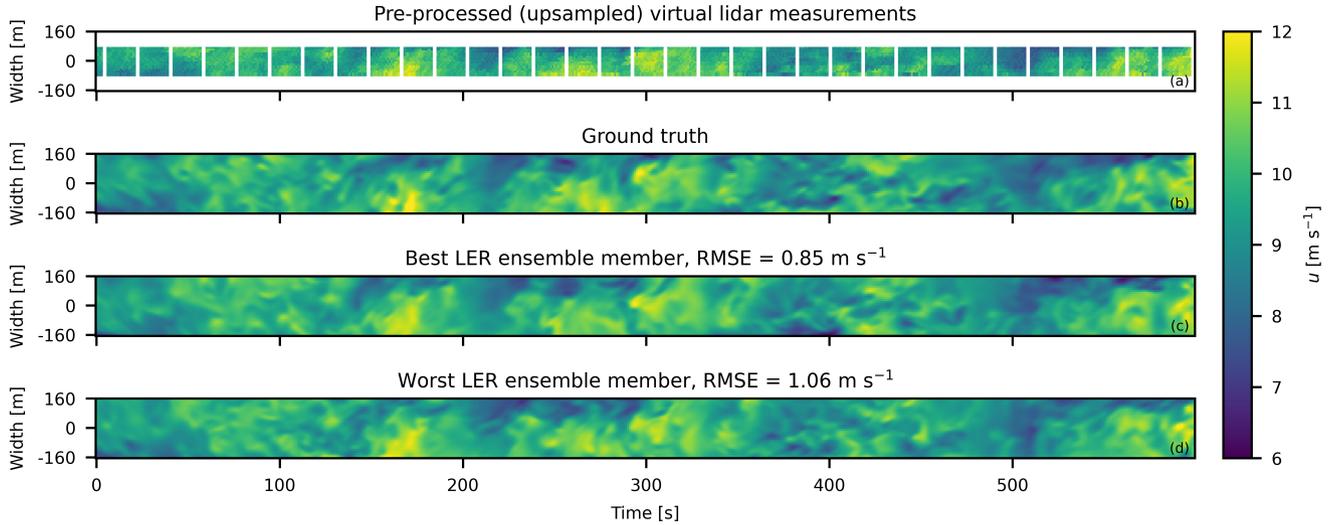

**FIGURE 5** A top-down view of $u$ at hub height of (a) noisy, upsampled measurements, (b) the ground-truth winds, (c) the best reconstruction based on root-mean-square error (RMSE), and (d) the worst reconstruction for OSSE3. The white, vertical stripes in (a) are the 2-s period during which the lidar finishes a sweep and resets to its initial position.

Using this domain boundary condition data, we then run three separate inflow-outflow simulations for the same time window. In each of these, we run an OpenFAST model of the RAAW turbine[6]. Each of the turbines has an $x$-coordinate of 7,680 m. The three turbines have $y$-coordinates of 3,840 m, 7,680, and 11,520 m, and we refer to these cases as OSSE1, OSSE2, and OSSE3, respectively. We collect nacelle-mounted synthetic lidar measurements akin to our real-world PPI scans for the last 704 s of the OSSEs. We corrupt these perfect lidar measurements of the flow field by using our estimated real-world corruption procedures (Section 3.3). We also collect the ground-truth LES inflow planes $3D$ upwind for the same time window. In doing so, we can reconstruct ensembles of inflow planes using the synthetic noisy lidar measurements, and we can compare all four field variables to ground-truth inflow planes $3D$ upwind. This comparison is not feasible with the RAAW campaign measurements, where instead we can only compare to line-of-sight SpinnerLidar measurements $1D$ upwind. In addition to the nacelle-mounted lidar measurements on each turbine, we also collect synthetic SpinnerLidar measurements. These SpinnerLidar measurements are not used in this section (which is about inflows at $3D$), but they are later used to give context to the real-world validation results (Section 5.4.2)

## 4.2 | Qualitative visualizations of the inflow planes from the OSSEs

Do our reconstructed turbulence fields match the ground-truth fields "well"? Unfortunately, there is no single metric that can be used to satisfactorily answer this question, and often, the performance of an atmospheric model is assessed by simultaneously examining multiple quantities of interest (e.g., Taylor (2001)[72], Chang and Hanna (2004)[73]). We use a combination of qualitative and quantitative assessments to provide responses to this question and suggest the usefulness of LER for utility-scale wind turbine inflow reconstruction. Due to space constraints within the manuscript, we only include a handful of visualizations here, and we include additional figures in the Supplementary Information (SI Section 1). There, we include figures for hub-height cross sections and time-height cross sections of all four variables. In this subsection, we limit the visualizations to the $u$ field for OSSE3.

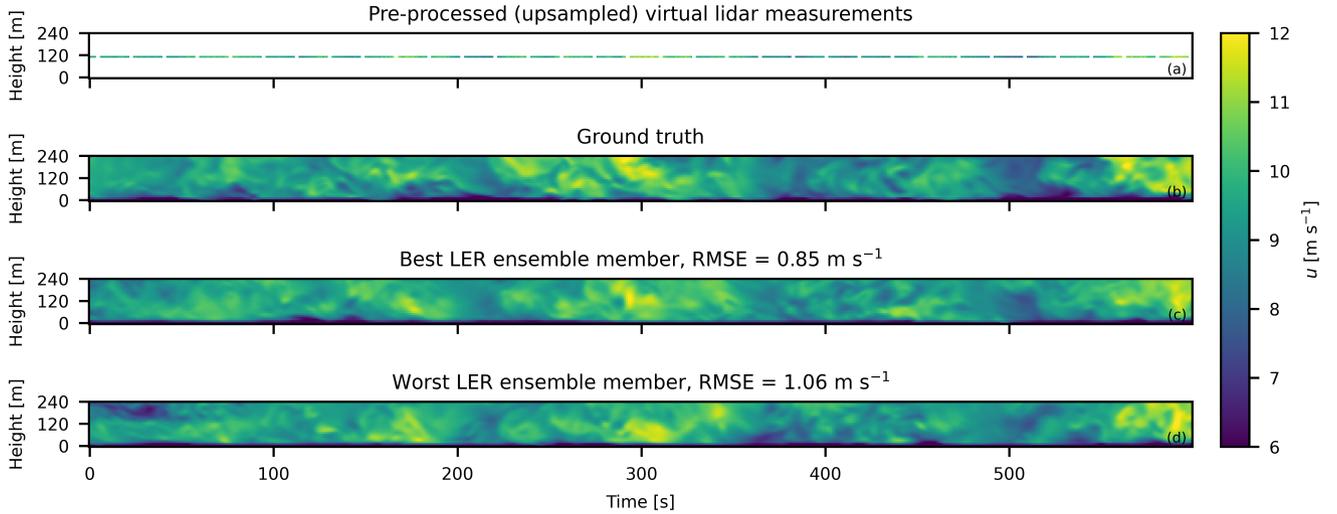

**FIGURE 6** Same as Fig. 5, except a side view through $y = 0$ m.

First, we qualitatively demonstrate a successful reconstruction by visualizing time-width cross sections of $u$ at hub height (Fig. 5). We compare the preprocessed virtual lidar observation, the ground-truth field where this lidar is measuring, and the best and worst LER ensemble members as determined by root-mean-square error across the entire ensemble. We suggest two key takeaways from this figure. First, the LER fields look stylistically similar to the ground-truth LES field. They have clusters, streaks, and patterns of wind behavior that look similar to the ground truth. Second, the LER fields track the hub-height observations on a second-by-second basis, and they extrapolate from the lidar measurements in a way that appears reasonable to the eye. For example, near 180 s the lidar winds are faster, and correspondingly, both the ground-truth winds and the reconstructed winds are faster. For comparison, the reconstructed $v$, $w$, and $T$ fields (SI Section 1) look stylistically similar to the ground-truth data, but they do not track second-by-second behavior of the ground truth as proficiently as for $u$.

Second, we demonstrate that our LER algorithm produces visually realistic $u$ fields above and below our input measurements by now visualizing time-height cross sections along the center ($y = 0$ m) of the lidar scan (Fig. 6). The reconstructions still qualitatively track the timing of the patterns visible in the ground truth, despite the limited vertical extent of the observations. The reconstruction fields also do not show any obvious discontinuities or other artifacts associated with the sequential sampling process (Section 3.4.2). We emphasize that these cross sections look stylistically different than those generated with a kinematic turbulence generator (e.g., Fig. 4 within Doubrawa et al. 2024[59]), which produce fields that visually have larger variability between neighboring grid cells. Both these qualitative comparisons of $u$ (as well as the supplemental comparisons for other variables) suggest that we are able to successfully track true second-by-second inflow to some capacity.

## 4.3 | Quantitative assessment of the inflow planes from the OSSEs

Next, we turn toward quantitative assessment techniques to demonstrate the capabilities of our LER algorithm. We begin by calculating the Pearson correlation coefficient[74] through time between the true inflow and the reconstructed inflows for each matching grid cell, and then we average over the crosswind dimension (Fig. 7). This metric can be used to assess how well two time series track one another, where 1 denotes perfect correlation, 0 denotes no correlation, and -1 denotes perfect anticorrelation. The strongest correlation values are found for hub-height values of $u$, which is where there are observations. The ensemble spread is also narrow at this location, denoting high reconstruction confidence. When averaged across the crosswind dimension, the ensemble mean of the hub-height correlation value is approximately 0.85. If only looking at the profile at $y = 0$ m (not shown) instead of the horizontally averaged profile, that correlation value is 0.93; conversely, at the horizontal edge of the reconstruction, that correlation value is 0.65. The correlation values for $u$ decay when moving vertically away from hub height, and the ensemble spread grows. Depending on the case study, ensemble mean $u$ correlation values at the top and bottom of the inflow can be as large as 0.5 or as low as 0.2. The nonzero values denote that the diffusion model is able to learn non-negligible time correlations for $u$ at unobserved heights. Similarly, while the diffusion model only receives observations of $u$, the Pearson correlation coefficient for $v$, $w$, and $T$ takes on positive values depending on the height and case study. These consistently positive values denote that the diffusion model similarly learns correlations between $u$ and the unobserved field variables.

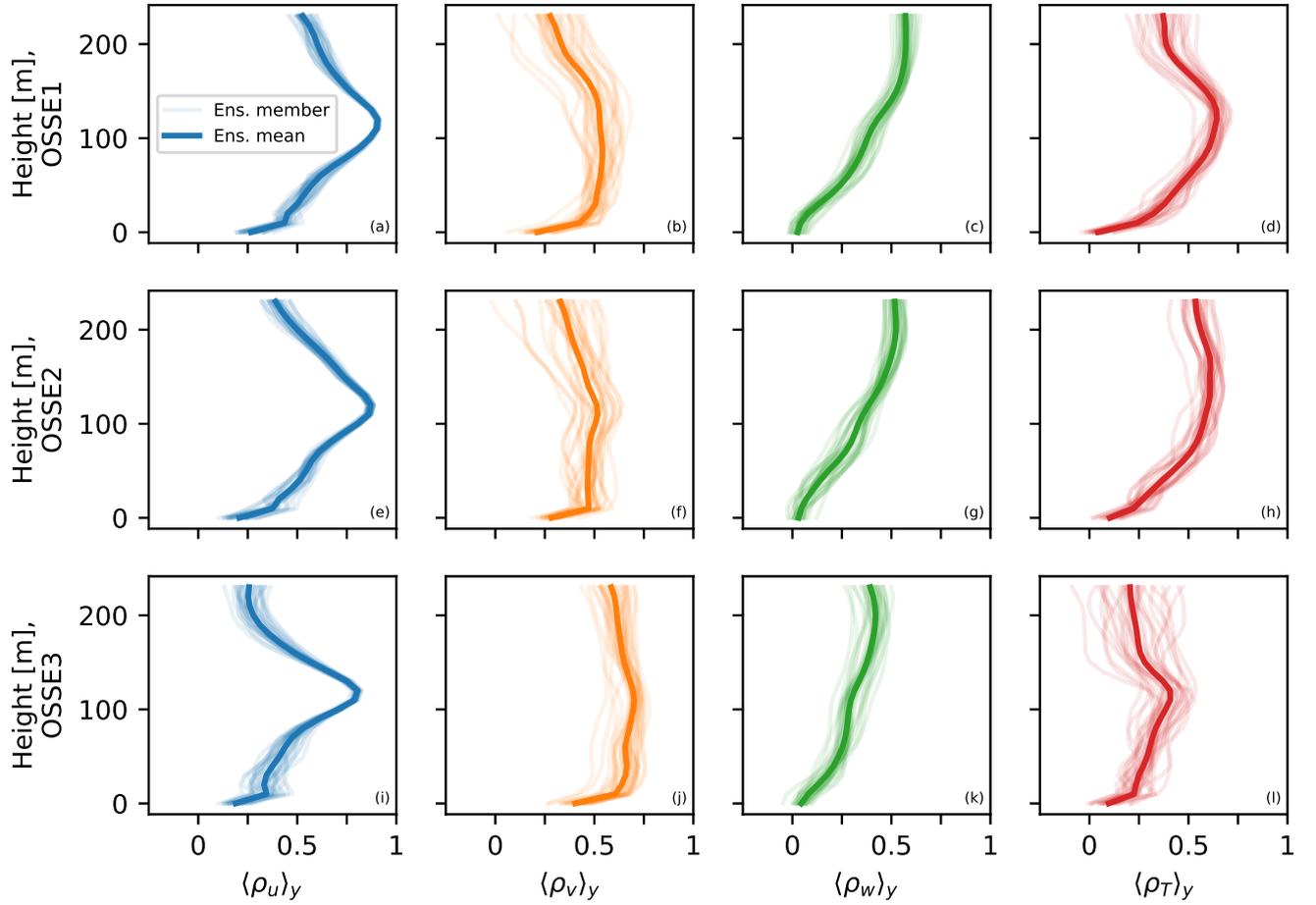

**FIGURE 7** Vertical profiles of Pearson correlation coefficients between the reconstructed inflows and the ground-truth inflow, averaged over the crosswind dimension.

Next, we compare reconstructions of OSSE inflows using more traditional metrics, namely, horizontally averaged vertical profiles of the mean and variance of fields taken along the time dimension (Fig. 8). Due to space constraints, we visualize these quantities only for OSSE3 within the manuscript, and we share the similar plots for OSSE1 and OSSE2 in the Supplementary Information (SI Section 2). In general, the mean profiles (Fig. 8 a–e) of the reconstructions match the shape of the ground-truth profiles, and the ensemble spread typically is wide enough to envelope the ground-truth profile. The reconstruction ensemble only fails to span the mean ground-truth values for horizontal wind speed and $u$. The largest two-dimensional wind speed deviations occur near rotor bottom. The deviations between the ground truth and the ensemble mean at these locations is no larger than 0.3 m s$^{-1}$ (or 3% relative error) across all three case studies. We believe these biases are driven by small differences between the atmosphere in the training dataset and the OSSE environment because of the presence of a turbine in each OSSE. Across all three OSSE case studies, the reconstructions slightly underpredict wind speed shear—in the ground truth, the wind speed measurements at the rotor tip are ∼0.6 m s$^{-1}$ stronger than they at the rotor bottom, whereas this difference is ∼0.2 m s$^{-1}$ for the ensemble mean of the reconstructions. Put differently, the presence of the turbine appears to increase the shear 3$D$ upwind of this turbine. This effect is counter to the findings of Forsting et al. (2018)[75], which suggest that for low shear inflow, the presence of the turbine should not change the magnitude of the shear, although this study is conducted at the rotor disk instead of 3$D$ upwind. We hypothesize that this slight underestimation in the reconstructed shear profile could simply be caused by random errors or possibly by small near-surface wind acceleration, as suggested by the discussion about TI below.

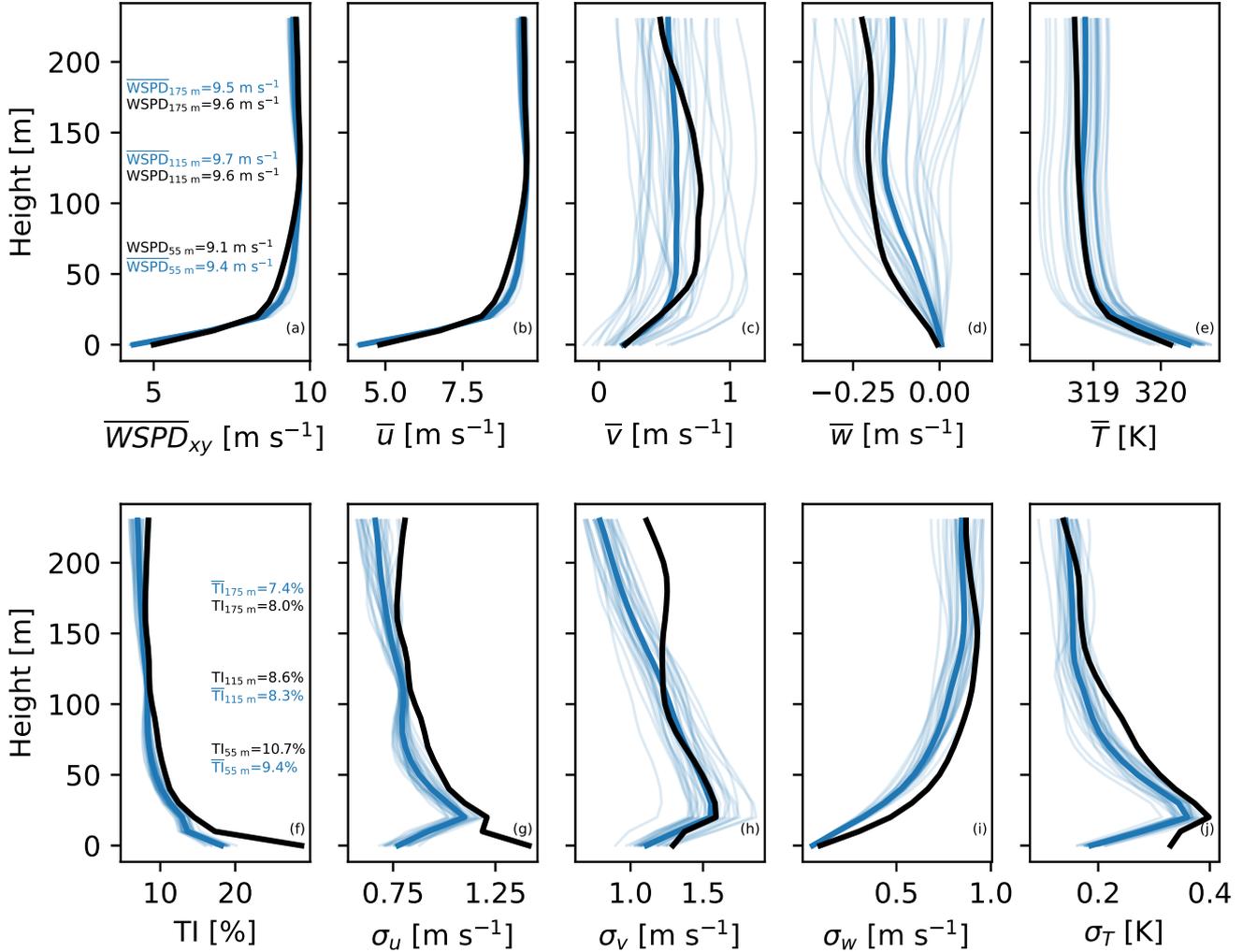

**FIGURE 8** (a–e) Mean horizontal wind speed, $u$, $v$, $w$, and $T$ profiles. Mean values for each ensemble member are calculated across both time and the lateral dimension. Horizontal wind speed values are stated near rotor bottom, hub height, and rotor top. (f–j) Turbulence intensity as well as standard deviations of $u$, $v$, $w$, $T$. Turbulence intensity for each ensemble member is calculated by taking the standard deviation and mean across time at each $y$ location and then taking the mean of those across $y$. Standard deviation profiles are calculated by taking standard deviations across both time and $y$. Turbulence intensity values are stated near rotor bottom, hub height, and rotor top.

The reconstructions generally match TI and standard deviations (Fig. 8 f–j), though they show larger errors than for the mean statistics. In general, second-order statistics are more difficult to accurately reconstruct than first-order statistics. Whereas mean ground-truth values were generally spanned by the ensemble, the same is not true for TI and standard deviations. In general, these quantities are underpredicted in the reconstructions. For example, $\sigma_u$ and $\sigma_v$ are underpredicted aloft for OSSE3, and $\sigma_w$ and $\sigma_T$ are underpredicted closer to the surface. We believe that, in general, the underprediction of these statistics is due the use of a 256-s-long window by the LER inpainting algorithm rather than a 10-min window. Put differently, the inpainting algorithm accurately learns the standard deviation of blocks of inflow that are 256 s long and thus misses the contribution from slower turbulent motions. During the development of our algorithm (not shown), we tested the use of shorter blocks that were 64 s long, and we similarly saw that the prototype algorithm produced 10-min-long inflows that had the variance associated with inflow that is 64 s long. We hypothesize that the near-surface underprediction of TI and $\sigma_u$ may be explained by small, near-surface flow acceleration[76] in the turbine simulations. The reconstructed inflows all have a TI of ~20% at the surface, which is consistent with the TI in the training dataset (Fig. 2). By comparing the training dataset (where there are no turbines) to the OSSE simulations, we see that the turbines substantially modify the near-surface TI. With these caveats about the TI near the ground, we find that TI profiles across the rotor disk are accurate to within approximately 1.3 percentage points across all cases. The reconstructed TI is most accurate near hub height.

# 5 | VALIDATION AND VERIFICATION AGAINST SPINNERLIDAR MEASUREMENTS

While the OSSE inflow study at $3D$ informs the accuracy of our LER approach under ideal conditions, to be useful for validating real-world turbine models, we need to demonstrate that the LER technique can accurately reconstruct inflows that occurred in the real world. To do this, we generate inflows using real-world nacelle-mounted lidar measurements for three periods of the RAAW campaign (Section 2.2). We validate against real-world SpinnerLidar measurements for the coincident periods. However, the reconstructed inflow planes at $3D$ are not directly comparable to the SpinnerLidar measurements at $1D$—they are located at different upstream distances from the turbine, the SpinnerLidar winds are subject to stronger induction, and the SpinnerLidar measures line-of-sight velocity along a spherical surface instead of providing $(u, v, w)$ data in a plane. In this section, we first describe the use of "reconstruction simulations" (Section 5.1) and SpinnerLidar data postprocessing (Section 5.2) that enable us to compare our reconstructed inflow to SpinnerLidar measurements. Then, we conduct qualitative (Section 5.3) and quantitative (Section 5.4.1) comparisons against the real-world SpinnerLidar measurements. In order to provide context to the real-world SpinnerLidar comparison results, we end this section by conducting additional SpinnerLidar analysis with the help of OSSEs (Section 5.4.2). In this section, we discuss simple similarities and differences between the observed and reconstructed SpinnerLidar data without diving into the mechanisms behind these results. In Section 6, we hypothesize processes that could drive the differences, and we tie the real-world validation results back to the inflow verification analysis (Section 4).

## 5.1 | Configuration of the reconstruction simulations

To enable us to compare our reconstructed inflow to the SpinnerLidar measurements, we use our ensemble of inflows as inflow boundary conditions to drive an ensemble of LES runs. We dub these simulations "reconstruction simulations". In these simulations, we situate the RAAW turbine $3D$ downwind of the inflow boundary, and we simulate a synthetic version of the SpinnerLidar. The simulated SpinnerLidar measurements are simplified approximations of the real-world SpinnerLidar measurements, for example, lacking sensor noise. The simulated SpinnerLidar matches the real-world scanning rate in time and space while accounting for probe volume averaging.

The reconstruction simulations are initialized with mean $u$ and $T$ profiles that match the observed profiles. Inspired by Troldborg et al. (2014)[77], we embed our turbulent inflow with spatial dimensions $(L_y, L_z) = (320\ m, 240\ m)$ within a laminar background flow that is sized $(640\ m, 1280\ m)$. We make this decision because we need to prescribe lateral boundary conditions and the upper surface boundary condition. By using a laminar buffer, we are able to prescribe periodic boundary conditions for the lateral boundaries and a free-shear boundary for the upper boundary condition while minimally impacting the SpinnerLidar region. As such, our reconstruction simulations are sized $(1280\ m, 640\ m, 1280\ m)$ using a 10-m grid resolution. Each simulation runs for 704 s, and we discard the first 104 s to mitigate transient behavior in the atmosphere and the turbine. To match possible dynamic induction effects that could impact inflow validation, our simulated turbines use a controller that synchronizes the modeled turbine's rotor position and blade pitch to the real-world turbine during the simulated time period. We note that all 90 of our reconstruction simulations that are driven by machine learning–generated turbulence (30 ensemble members across 3 cases) ran to completion without any obvious numerical instabilities, as suggested by the visual quality of SpinnerLidar reconstructions (Section 5.3).

While our simulations of SpinnerLidar measurements enable us to validate our reconstructed inflows, these simulations come with a few important caveats. As a first caveat, during normal turbine operation, the rotor disk changes its yaw position to stay aligned with the direction of incoming wind. Our inflow reconstruction approach computes the inflow in the reference frame of the rotor disk instead of a global coordinate system that is relative to, e.g., the cardinal directions. In other words, we assume a fixed yaw position and we conduct our reconstruction simulations accordingly. Second, we trained a single set of networks for one bulk atmospheric state, but in the real world validation we seek to match the behavior of three distinct 10-min periods. The three observational periods are similar, but they have slightly different bulk characteristics from each other as well as the 75-min-long period that was used to calculate the statistics for the data generation LES (Table 1).

## 5.2 | Postprocessing SpinnerLidar data

We postprocess both the real-world and reconstructed SpinnerLidar measurements to make them more amenable for analysis. We downsample the raw measurements from an unstructured grid at high temporal resolution into gridded bins that are 5 s long and that have a height and width of 10 m. We select these dimensions to minimize the data gaps on the structured grid while retaining a large degree of spatiotemporal detail, though we note that a few data gaps are still present with this binning strategy (e.g., the white cell in Fig. 10b).

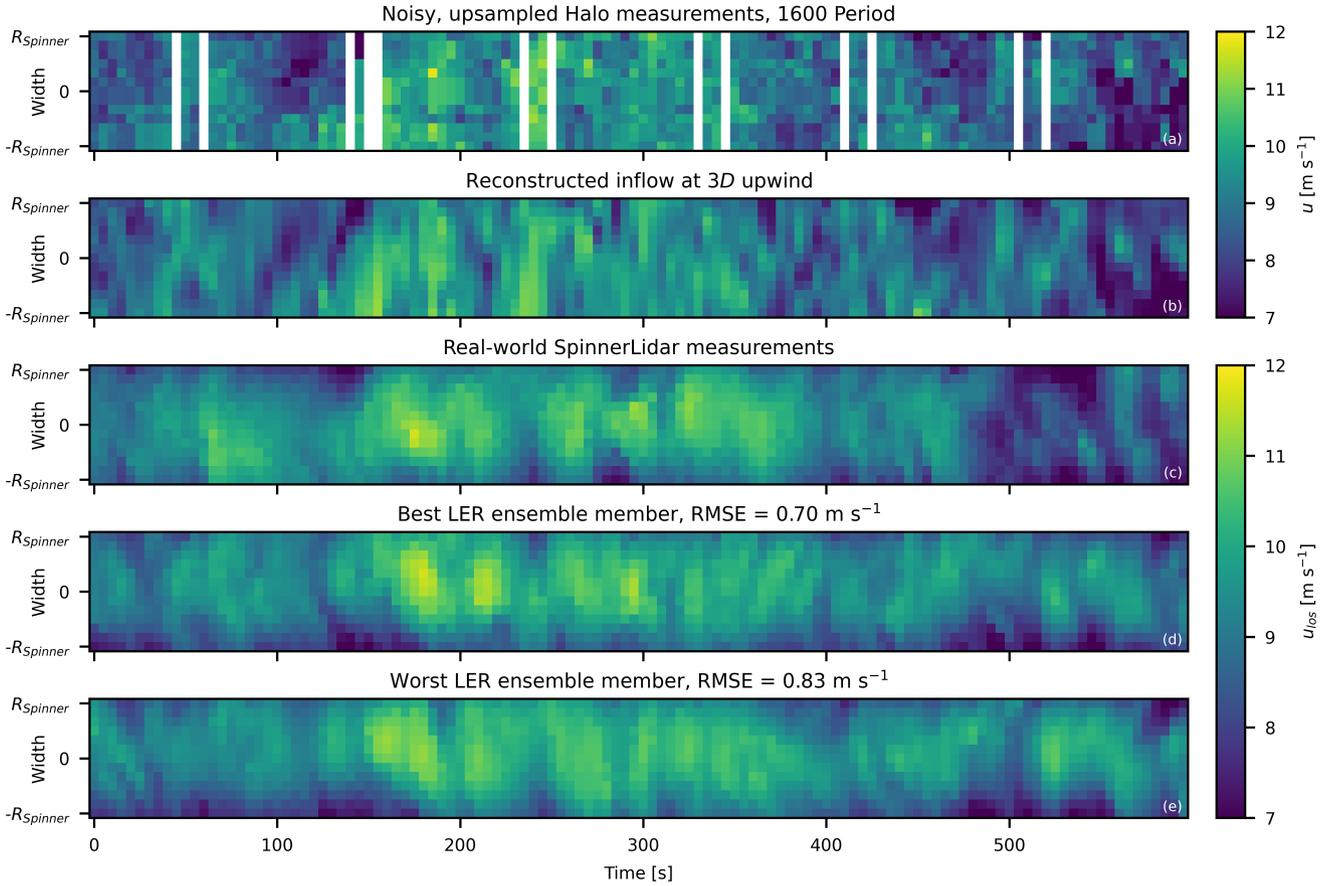

**FIGURE 9** A hub-height top-down view of (a) upsampled streamwise winds from the nacelle-mounted lidar, (b) the reconstructed inflow at $3D$, and (c) line-of-sight velocity from the real-world SpinnerLidar, as well as the (d) best and (e) worst line-of-sight reconstructions of SpinnerLidar measurements for the 1600 Period.

## 5.3 | Qualitative visualizations for the real-world SpinnerLidar comparison

Just like with the OSSE study, we characterize the quality of our real-world inflow by using a combination of qualitative and quantitative measures. We visualize the inflow at the SpinnerLidar scan via a top-down view (Fig. 9) at hub height and a side view (Fig. 10) through the $y = 0$ m centerline for one of the case studies in the body of the manuscript, and we include visualizations of the other periods in the Supplementary Information (Section SI 3). As with the OSSE inflow study, the reconstructed SpinnerLidar measurements stylistically look like the real-world SpinnerLidar measurements. The reconstructions track many of the turbulent speedups and slowdowns, even in the unobserved locations above and below the nacelle-mounted lidar. However, unlike in the OSSE study, the reconstructions occasionally miss some features that appear in the observations, such as the observed slowdown for the 1600 Period near 550 s. We discuss potential reasons for the mismatch between observations and reconstructions in Section 6.

We additionally visualize line-of-sight velocity time series at the rotor top height, hub height, and rotor bottom height to assess second-by-second tracking ability and confidence at a few key points (Fig. 11). As expected, the reconstruction ensemble mean tracks the measured winds the best at hub height, and the ensemble spread among different realizations is also narrowest at this height. Above and below, the reconstruction ensemble mean typically tracks with observations, though less frequently, and the ensemble spread is larger. At the unobserved heights, the reconstructions sometimes capture quick $u$ ramps (e.g., Fig. 11a at 375 s) but miss them at other times (e.g., Fig. 11c at 75 s). At hub height however, these wind ramps are typically accurately reconstructed. At hub height, the mean ensemble spread between the minimum and maximum reconstructed wind speed was 1.2 m s$^{-1}$, and across all three cases, the ground truth fell within the spread 76% of the time. At rotor bottom and rotor top, the ensemble spread between minimum and maximum reconstructed wind speed was approximately 1.7 m s$^{-1}$. The ensemble spanned the ground truth approximately 69.5% of the time at rotor top and 82.5% of the time at rotor bottom. Thus, the reconstructions generally span observations, though observations can show stronger second-to-second variability at times and move outside the spread of the ensemble.

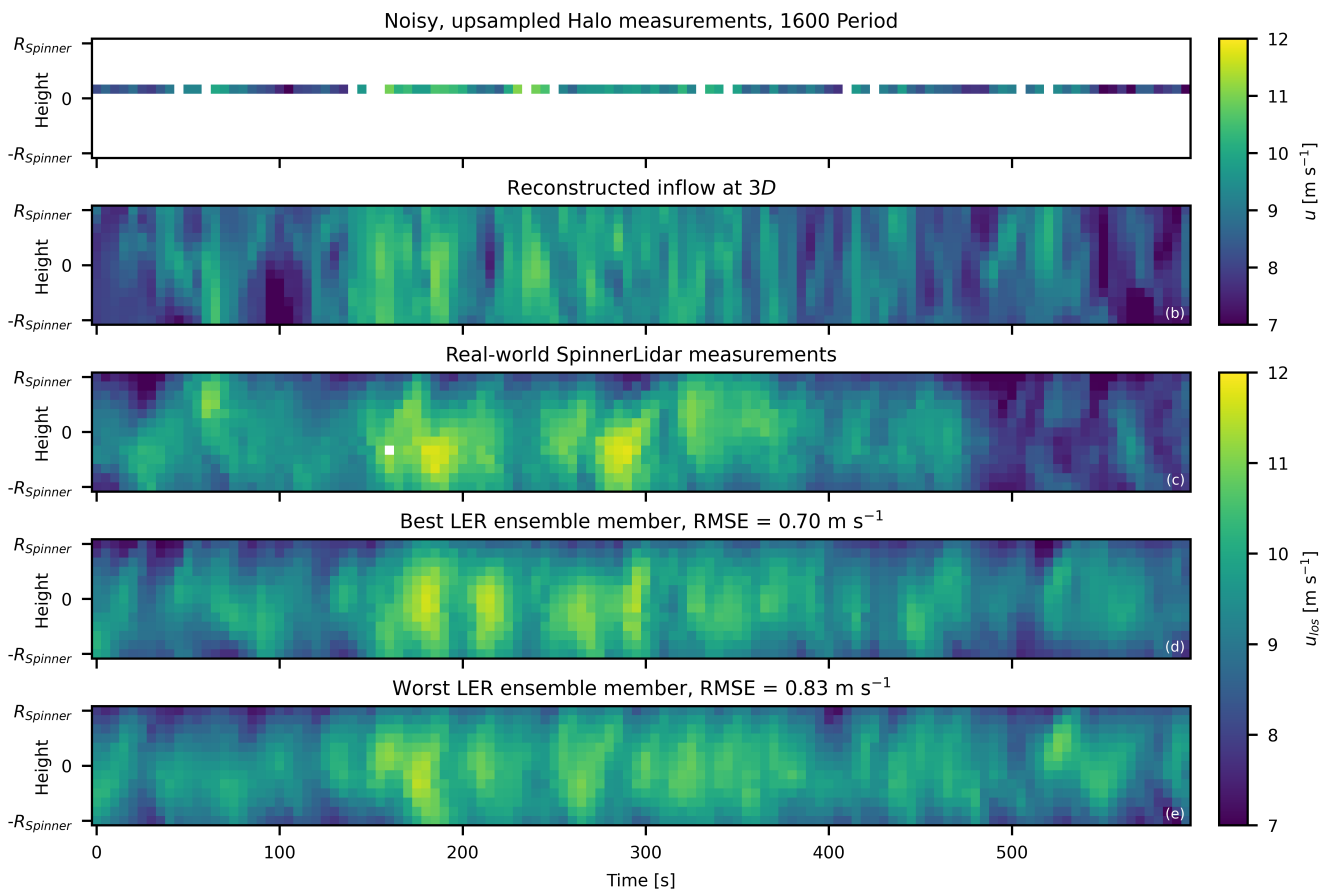

**FIGURE 10** Same as Fig. 9, except a side view through $y = 0$ m. The white pixel in (c) is present because of a lack of SpinnerLidar data for that bin in space and time.

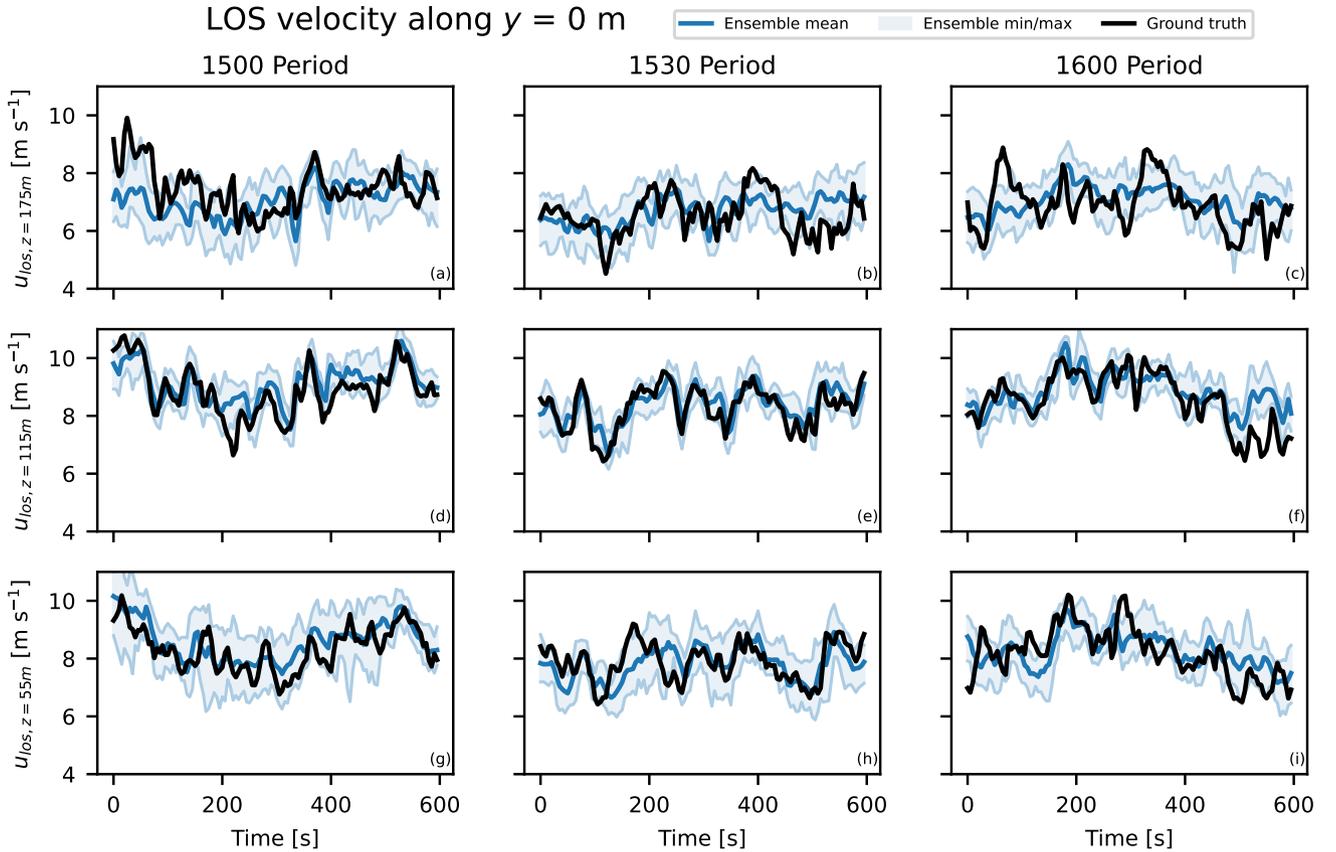

**FIGURE 11** Measured and reconstructed line-of-sight SpinnerLidar velocities at the top (row 1), center (row 2), and bottom (row 3) of the rotor disk at $y = 0$ m for (left column) the 1500 Period, (center column) the 1530 Period, and (right column) the 1600 Period.

These qualitative reconstructions also demonstrate that our diffusion models do not show an obvious problem with domain translation[78]. Machine learning algorithms are often trained on data from one distribution and then applied to data from a different distribution, and they can make nonsensical predictions in the new environment. Our lidar denoising algorithm is trained on noisy lidar measurements from LES, and it is applied to real-world lidar measurements. Our training data looks similar to the real-world measurements but is different. However, our LER algorithm appears robust to this shift.

## 5.4 | Quantitative assessment against SpinnerLidar measurements

### 5.4.1 | Validation against real-world SpinnerLidar measurements

When comparing line-of-sight velocities across the entire (time, width, height) box of SpinnerLidar data at once, the reconstructed winds show little bias (Fig. 12). We subtract the real-world SpinnerLidar data from the reconstructed SpinnerLidar data at each grid cell for the three periods and aggregate the errors into a histogram. The mean error is small: 0.08 m s$^{-1}$ for the 1500 Period, 0.01 m s$^{-1}$ for the 1530 Period, and 0.11 m s$^{-1}$ for the 1600 Period. The standard deviations for the respective periods are respectively 1.08 m s$^{-1}$, 0.82 m s$^{-1}$, and 0.87 m s$^{-1}$.

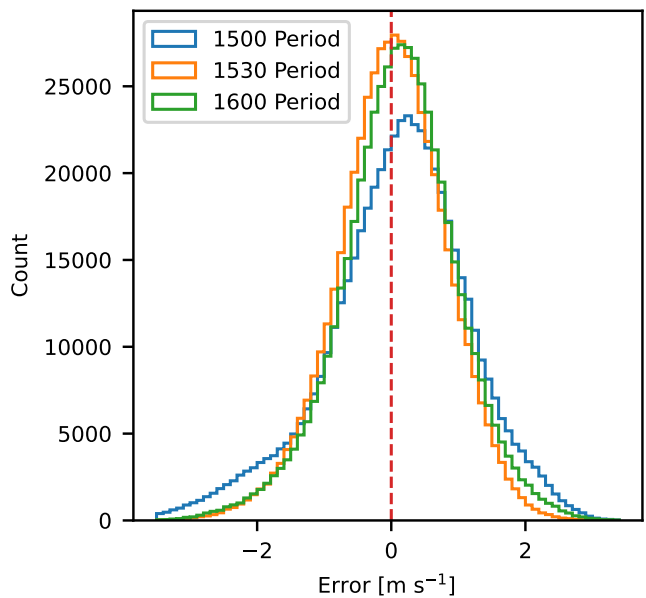

**FIGURE 12** Reconstructed inflow minus observed inflow at each grid cell for gridded SpinnerLidar data for each of the three periods of interest.

The reconstructed SpinnerLidar measurements produce power spectra that match desired behavior (Fig. 13). We calculate the power spectra for hub-height data across the width of the SpinnerLidar for all three real-world periods. Due to the small number of time stamps ($n = 120$), spectra are noisy at each $y$ location, so we average all spectra across the $y$ dimension, and we furthermore calculate the ensemble average spectrum across all the reconstructions. We also plot the ensemble maximum, and we omit the ensemble minimum because near-zero power spectrum values would render the plots illegible. At all heights, both the measured spectra and the reconstructed spectra approximately follow a -5/3 decay. The spectra of the reconstructions do not show an obvious deficiency at either the small spatial scales or the large spatial scales. Machine learning algorithms can struggle to learn variability on smaller spatial scales due to a machine learning spectral bias[79], and our previous initial condition reconstruction algorithm suffered from this problem[55]. We hypothesize that we avoided the spectral bias here due to the use of a simpler architecture here that lacks a variational autoencoder as well as improved model assessment infrastructure to characterize the quality of machine learning output during training. The ensemble average of the reconstructed spectra is smaller than the observed spectra, which is consistent with the underestimated TI at the SpinnerLidar (discussed more below).

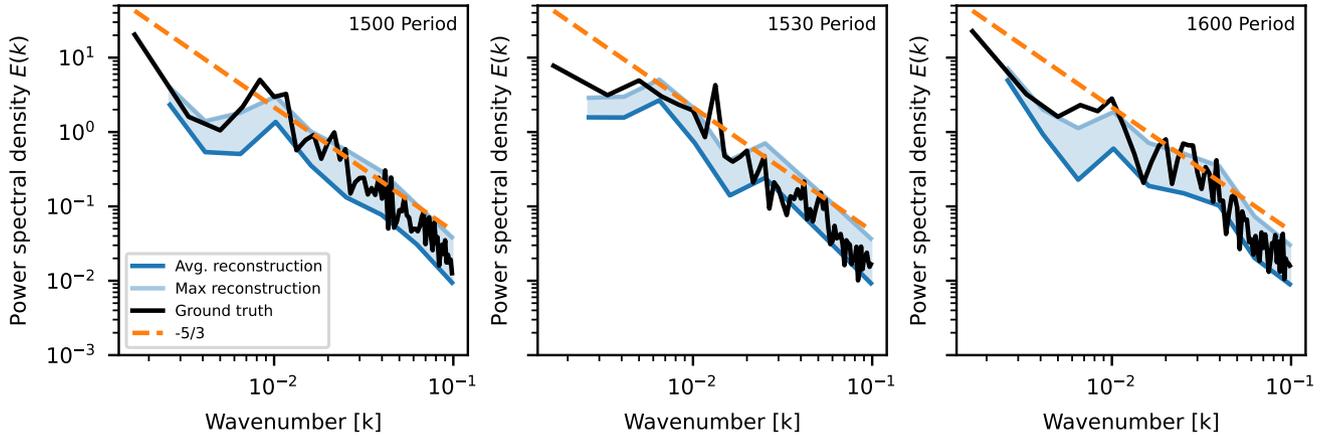

**FIGURE 13** Observed and modeled hub-height power spectra that have been averaged across *y* for (a) the 1500 Period, (b) the 1530 Period, and (c) the 1600 Period.

The reconstructed SpinnerLidar measurements approximately track the second-by-second real-world behavior, as quantified by the Pearson correlation coefficient along the *y* = 0 m centerline (Fig. 14 a,d,g). As was the case in the OSSE inflow study (Section 4.3), correlation is strongest at hub height (∼0.75) and it tapers off with increasing distance from this location (∼0.25 at rotor top, ∼0.5 at rotor bottom).

The reconstructed time-averaged winds match the observations better for some cases and match more poorly for other cases (Fig. 14 b, e, h). In the 1530 Period, the ensemble mean of the modeled winds matches the measurements at all heights to within a value of 0.18 m s$^{-1}$. In this period, the shape of the observed and modeled profiles closely match as well. In contrast, in the 1500 Period the modeled winds overpredict by a maximum of 0.40 m s$^{-1}$, and the shapes of the modeled and observed profiles disagree. With respect to the profile shape, we note that the shape of these mean profiles is curved because we are comparing line-of-sight values from the SpinnerLidar as opposed to streamwise velocities.

Just as with the velocities, the accuracy of the line-of-sight TI reconstruction varies across periods (Fig. 14 c, f, i). We calculate the line-of-sight TI by taking the standard deviation of line-of-sight wind speed through time and dividing that value by the time-averaged line-of-sight wind speed. In the 1530 Period, the time-averaged mean profile was most accurate, and for this period, the TI profile is also the most accurate. While the 1500 Period shows the greatest deviation in the mean wind speed, the 1600 Period shows the largest deviation in the TI profile. In this period, the ensemble mean profile underpredicts on the order of 5 percentage points, and the observed profile falls outside of the ensemble spread at all heights. We note that this error in TI for the 1600 Period is related to the reconstruction error in the last 100 s (Figs. 10 and 11f). If we recalculate TI omitting this window (not shown), the reconstruction matches the true TI substantially better. This error in TI reconstruction motivates the work in the following subsection.

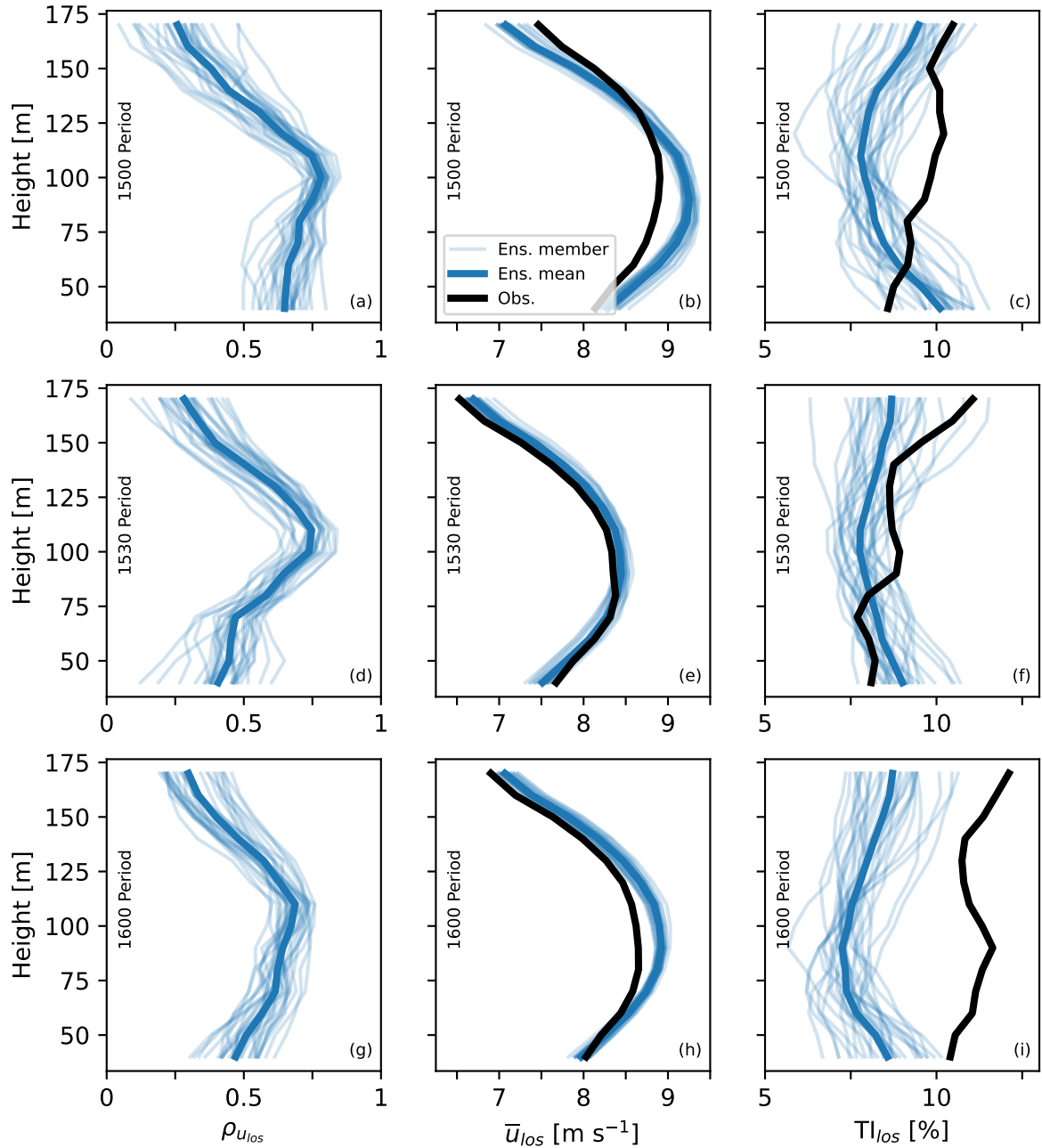

**FIGURE 14** Vertical profiles along the centerline $y = 0$ m of the (left) Pearson correlation coefficient, (center) 10-min averaged line-of-sight velocity, and (right) standard deviation of line-of-sight velocity over 10 min for the three real-world periods considered.

## 5.4.2 | Verification of SpinnerLidar reconstructions using OSSEs

To give context about potential sources of error (Section 6) in the real-world SpinnerLidar validation analysis, we return to the OSSEs from before. In addition to collecting nacelle-mounted lidar measurements in each of the OSSE case studies, we also collected SpinnerLidar measurements (Section 4.1). Here, we use the inflows that were reconstructed at $3D$ in the OSSE studies to drive a new ensemble of reconstruction simulations that also include synthetic SpinnerLidar measurements. Then, just like in the real-world SpinnerLidar study, we can compare our reconstructed SpinnerLidar measurements to ground-truth SpinnerLidar measurements. However, in this section, the ground-truth data come from the OSSEs. By running this OSSE comparison, we can address the question "How well can we reconstruct SpinnerLidar observations under ideal conditions?" Using these results, we calculate profiles of the $y = 0$ m Pearson correlation coefficient, time-averaged line-of-sight velocity, and the corresponding TI (Fig. 15). We compare these results to the real-world SpinnerLidar validation results as well as the OSSE inflow verification results in Section 6.2.

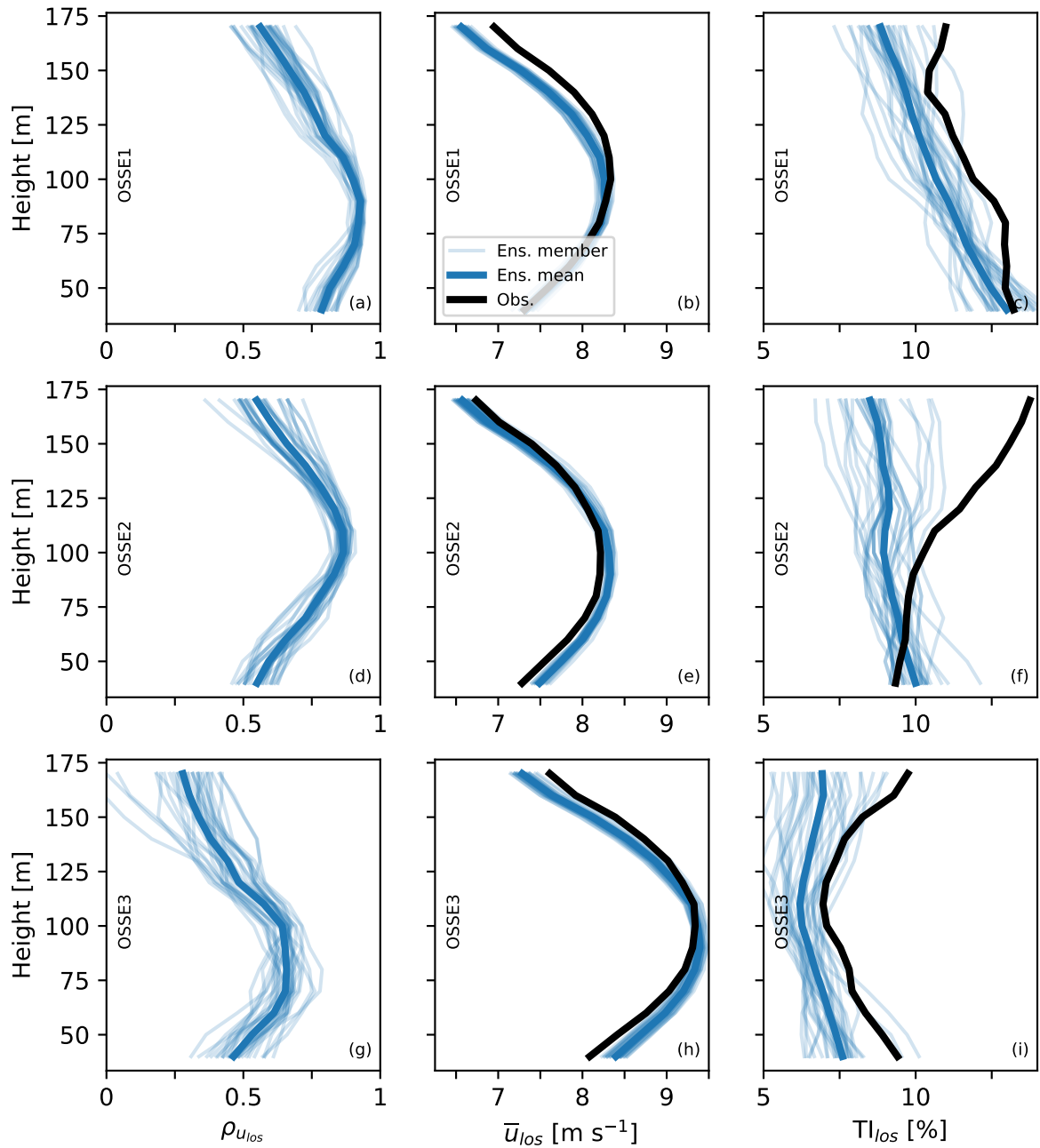

**FIGURE 15** Same as Figure 14, except for the OSSE SpinnerLidar study.

# 6 | SOURCES OF ERROR AND DISCUSSION ON LER ACCURACY

## 6.1 | Sources of OSSE and real-world reconstruction error

To develop our inflow reconstruction methodology that blends observations with an LES model of the atmosphere, we made several assumptions and approximations. These factors contribute to reconstruction errors, particularly for real-world atmospheres. We document questions associated with these assumptions below in order to inform our discussion on LER accuracy as well as for avenues of potential future investigation.

Regarding the preprocessing of real-world inflow Halo measurements (Section 3.2):

- What are the errors associated with the cosine projection, especially pertaining to measurements that are not time-averaged?
- What are the errors associated with temporal upsampling with our choice of PPI scan design, particularly when accounting for nearest-neighbor interpolation to account for blade passage?
- What are the errors associated with regridding the measurements from a polar coordinate system to a rectangular one?
- How significantly is the temporal upsampling process affected by small rotor yaw maneuvers in the real world (Section 5.1)?

Regarding the generation of the training dataset via LES (Section 3.3):

- In general, how accurately can an LES code model the dynamics of a real-world ABL?
- How accurately does our particular LES run for the training dataset (i.e., doubly periodic boundary conditions, steady-state bulk forcing) model the dynamical state of our specific observational period of interest? We examined time-averaged wind speed and TI (Fig. 2), but are there other important quantities of interest that we should match for a turbine inflow?
- How closely does our approximated lidar corruption process match the real-world lidar corruption process?
- In general, we wish to apply our algorithm when a turbine rotor is spinning, but we neglect turbine induction effects in the training dataset. How large is the error associated with neglecting induction?

Regarding the machine learning algorithms (Section 3.4):

- The diffusion models learn to sample from a conditional probability density function associated with the training dataset, enabling stochastic sampling. How accurately does it learn that probability density function?
- Were the models trained on a sufficient amount of data for our period of interest?
- How significant are the errors associated with the autoregressive (sequential) inflow generation process?

Regarding the application of our LER algorithm for validation against SpinnerLidar measurements (Section 5):

- What is the error associated with neglecting turbine yaw maneuvers during validation?
- We trained our algorithm on data for one set of atmospheric forcing, but we apply it in three environments with three slightly different sets of forcing. How large is the error associated with this mismatch?
- We only validate our algorithm during three 10-min blocks. How would our validation results change if we validated against more time windows with a comparable atmospheric state?
- What is the error associated with the use of our reconstruction simulations in order to propagate reconstructed inflow data to the SpinnerLidar region?

## 6.2 | Synthesizing real-world SpinnerLidar validation results, OSSE SpinnerLidar verification results, and OSSE inflow results

We have thus far assessed the skill of our LER technique using a combination of qualitative and quantitative methods. The qualitative analyses suggest that—across three synthetic case studies—our LER technique consistently reconstructs $u$ in a way that seems reasonable by eye, especially at hub height. This same qualitative behavior is generally true for real-world SpinnerLidar line-of-sight velocity, though the reconstructions did inaccurately reconstruct hub-height behavior over a 100-s period in Fig. 9. A similar but shorter mismatch occurred in the 1530 Period (SI Section 3), but the 1500 Period does not show any substantial qualitative mismatches. Crucially, the qualitative results demonstrate that our reconstruction simulations often track second-by-second wind dynamics.

The quantitative analyses give more details on the performance of our LER technique. From the inflow verification study, we find that our LER technique tracks the observed $u$ on a second-by-second basis fairly well, and it tracks unobserved $u$ as well as unobserved $v$, $w$, and $T$ with some lesser skill. This verification analysis also suggests that our inflows see some small impacts on mean quantities due to the presence of a turbine, and the second-order statistics (as well as TI) are additionally impacted by our autoregressive inflow generation process that uses time windows shorter than 10 min.

We next discuss SpinnerLidar validation (Section 5.4.1) and verification (Section 5.4.2) of averaged line-of-sight velocity. The reconstructions in the OSSE verification show smaller bias in the time-averaged line-of-sight velocity than they do in the real-world validation study. Indeed, looking across the three OSSE case studies at all the heights, reconstructions look relatively unbiased in the verification study, though some small bias between the reconstructions and the true winds may be present if looking at just one study. This result suggests that it is acceptable to use reconstruction simulations in order to propagate inflow data at $3D$ to the location of the SpinnerLidar. The relatively unbiased verification results also suggest that the error in the real-world, time-averaged line-of-sight velocity stems from a mismatch between our synthetic OSSE testing environment and the real world. These mismatches in the means could stem from several sources: differences between the simulated ABL dynamics in the training dataset and the real-world ABL, differences between the simulated turbine induction and real-world induction, and differences between the simulated turbine operation (e.g., yaw) and real-world turbine operation.

Regarding reconstruction of TI, in two SpinnerLidar verification case studies (OSSE1 and OSSE3), the shape of the TI profile is accurately reconstructed. However, the TI is slightly underestimated in these reconstructions ($\sim$1–2 percentage points), which is consistent with the underprediction of TI at $3D$ that at least partially stems from the use of the autoregressive inflow generation process. Notably, in OSSE2, the TI reconstructions (Fig. 15f) show large error ($\sim$5 percentage points). At the same time, the TI at $3D$ for this same case across the rotor disk was substantially more accurate with a maximum TI error of 1.1 percentage points (SI Section 2), which is comparable to the accuracy for the other two OSSEs at the inflow. However, we note that the TI at the inflow in these figures was calculated by averaging across the streamwise dimension, whereas they are visualized at $y = 0$ m for the SpinnerLidar, which means that these TI values strictly are not comparable. The behavior of TI in OSSE2 compared to that in OSSE1 and OSSE3 leads us to hypothesize that TI is a metric that is particularly sensitive to random errors, and to validate against it, it is particularly important to look across a large number of time windows. Relating these synthetic results back to the real world, we note that this error in the TI of OSSE2 of $\sim$7 percentage points is comparable to the $\sim$5 percentage point error observed in the real-world validation during the 1600 Period (Fig. 14i). Thus, it is likely that the TI validation during the 1600 Period is also subject to the same sensitivity to random errors.

# 7 | CONCLUSION

In this paper, we develop an approach to reconstruct turbine inflows by blending observations with data from large-eddy simulation (LES) runs using a machine learning algorithm. We state that our methodology addresses the "large-eddy reconstruction" (LER) problem to distinguish from the problem of model-observation blending for coarser spatiotemporal time scales employed by mesoscale-microscale coupling techniques. We built trust in our LER technique by running a synthetic verification study based off a virtual field campaign and also by validating against measurements from the RAAW field campaign.

We summarize the key takeaways from our synthetic verification study:

- Our LER technique reconstructs inflow data ($u$, $v$, $w$, $T$) at $3D$ upwind of a turbine that visually is stylistically similar to true inflow data.
- Our technique matches the second-by-second behavior of the true (synthetic) fields best where it has access to lidar measurements of $u$, and it extrapolates $u$ regions without observations in a plausible manner.
- The reconstruction algorithm learns some amount of physical correlation between $u$, $v$, $w$, and $T$.
- While the reconstructions generally matched the (synthetic) ground-truth field, there were deviations at times. These deviations likely stem from (1) differences between the algorithm training environment and the virtual field campaigns (e.g., turbine induction) and (2) hardware limitations and possible limitations associated with our specific machine learning algorithm architecture.

We similarly summarize conclusions from our real-world validation study:

- Our machine learning–generated inflows successfully interfaced with our LES code as boundary condition data across 90 simulations, and we did not see any obvious numerical instabilities.
- Our inflows produced simulated SpinnerLidar measurements that stylistically looked like real-world SpinnerLidar measurements.
- Our simulated measurements were able to track real-world, second-by-second measurements with some degree of skill. Across the ensemble of reconstructions, centerline mean Pearson correlation coefficients varied between 0.25 and 0.75, depending on the height.
- Our simulated measurements matched the power spectra of the real-world observations and the -5/3 decay law.
- Depending on the case study, we showed varied skill in being able to accurately reconstruct 10-min average line-of-sight velocity. We match time-averaged velocity profiles to within $\sim 0.18$ m s$^{-1}$ in our best case and to within $\sim 0.40$ m s$^{-1}$ in our worst case, depending on the height.
- We demonstrated that the accuracy for reconstructing turbulence intensity is similarly variable though less skillful.
- As with the synthetic verification study, we hypothesize reconstruction deviations stem primarily from (1) differences between the training environment and the real world and (2) algorithmic limitations.

In summary, for the first time, we demonstrate that LES can be used to reconstruct the second-by-second, three-dimensional state (time, width, height) of a real-world ABL on the scale of large, turbulent eddies. While we do not validate against four-dimensional observations (time, length, width, height), we hypothesize that our boundary conditions can also be used in tandem with an LES code to reconstruct the time-varying volumetric state of an ABL with some skill as well. While we tailor our algorithm for a wind energy application here, we envision that a similar methodology could be used for other kinds of field campaigns in the ABL (e.g., reconstructing trace gas concentrations, the state of a wildland fire, or clouds). We also, for the first time, generate observation-informed turbine inflows using observations in one region of space and we validate them against independent observations in a different region of space.

While we believe that our LER technique is useful for turbine validation case studies, we stress that this study focuses on the development and demonstration of a novel concept. Our inflow reconstruction comes with several limitations that would prevent its widespread use in the wind energy sector today. First, our algorithm is trained on a dataset with one specific set of hub-height wind speed, shear, turbulence intensity, and surface heat flux forcing. To reconstruct an atmosphere with different forcing, a new LES dataset must be generated, and the algorithm must be retrained. While this process may be feasible for a handful of case studies, it does not generalize as naturally as inflow reconstruction methods that use spectral-based models of the atmosphere—methods with which a user can request characteristics such as a desired shear and turbulence intensity. However, with the current machine learning model paradigm of training general-purpose foundation models and then fine-tuning those models, we could foresee a path forward to generalize our LER technique across multiple atmospheric states, especially as GPU hardware and LES codes continue to advance. Such an algorithm would also likely enable us to simulate inflows during complicated, non-stationary conditions. Second, our algorithm is trained specifically to use nacelle-mounted scanning lidar measurements that follow a specific scanning pattern, and it cannot handle input from other observations. This challenge could potentially be addressed by training a supervised machine learning algorithm to handle a wider source of inputs, or by adopting a machine learning architecture that is better suited to solving general-purpose inverse problems[80]. Third, given our hardware and our choice of using a U-Net based diffusion model, it is impractical to generate inflows that either have a larger spatial extent or more spatial resolution. This challenge could be potentially overcome by training more cascading networks (e.g., how we sequentially grow outward from the lidar measurements) or by adopting newer diffusion model architectures that scale well to larger samples[81].

We focused on developing and characterizing our LER technique in this paper, and we envision several lines of future inquiry after this work. To assess the accuracy of our inflow relative to those from other inflow generation techniques, we plan to participate in an inflow reconstruction benchmarking task that will be conducted under the International Energy Agency Wind Technology Collaboration Programme (IEA Wind Task 57). Through this benchmark, we plan to look at the real-world time periods studied in this paper as well as a new period with a different, more complex atmospheric state. While we do not have immediate plans to participate in IEA Wind Task 47, which is assessing aerodynamics models in turbulent conditions (building on the large intercomparison study of Boorsma et al. (2023)[82]), we believe our inflow reconstruction methodology could be useful there. Additionally, while the present study focused on atmospheric validation, we plan to use our inflows in upcoming aeroelastic turbine model validation studies. In particular, we are interested in applying our inflows to study the effect of turbine model fidelity on turbine model accuracy. Because we have demonstrated compatibility with an LES code, we can use the same LER inflow data to directly drive stand-alone aeroelastic simulations (e.g., OpenFAST), actuator disk and actuator line simulations within an LES code, and hybrid Reynolds-averaged Navier Stokes–LES blade-resolved simulations that employ fluid-structure interaction capabilities.

## AUTHOR CONTRIBUTIONS

**AR:** Conceptualization, Methodology, Software, Validation, Investigation, Writing - Original Draft, Writing - Review & Editing; **LAMT:** Conceptualization, Investigation, Supervision; **SL:** Conceptualization, Investigation, Writing - Review & Editing; **NH:** Conceptualization, Investigation, Writing - Review & Editing; **AS:** Investigation; **EM:** Investigation; **CI:** Data Curation; **DRH:** Conceptualization, Methodology, Writing - Review **TGH:** Conceptualization, Methodology; **NBdV:** Software; **PD:** Conceptualization, Investigation, Supervision, Project Administration, Funding Acquisition, Data Curation, Writing - Review & Editing

## ACKNOWLEDGMENTS

We thank the editor as well as the reviewers for their contributions to this work. This work was authored in part by the National Renewable Energy Laboratory, operated by Alliance for Sustainable Energy, LLC, for the U.S. Department of Energy (DOE) under Contract No. DE-AC36-08GO28308. Funding provided by the U.S. Department of Energy Office of Energy Efficiency and Renewable Energy Wind Energy Technologies Office. Support for the work was also provided by GE Vernova under CRADA 21-18140. The views expressed in the article do not necessarily represent the views of the DOE or the U.S. Government. The U.S. Government retains and the publisher, by accepting the article for publication, acknowledges that the U.S. Government retains a nonexclusive, paid-up, irrevocable, worldwide license to publish or reproduce the published form of this work, or allow others to do so, for U.S. Government purposes. The research was performed using computational resources sponsored by the Department of Energy's Office of Energy Efficiency and Renewable Energy and located at the National Renewable Energy Laboratory.

## FINANCIAL DISCLOSURE
None reported.

## DATA AVAILABILITY
Prior to final publication, we plan to store much of the data as well as the code used in this study in a publicly accessible location for long term storage.

## CONFLICT OF INTEREST
The authors declare no potential conflict of interests.

## SUPPORTING INFORMATION
Additional supporting information may be found in the online version of the article at the publishers website.

## APPENDIX

### A  DETAILS OF MACHINE LEARNING NETWORKS

#### A.1  Details of the lidar denoising network

We train one conditional diffusion model to use as part of the lidar denoising step of the LER pipeline. The model outputs uncorrupted $u$ data sized ($N_{time}, N_y$). The model is trained with a batch size of 32. We use a U-Net architecture for our diffusion model based off Palette[65], which is used for generic image-to-image translation tasks. We develop our code leveraging an unofficial PyTorch[83] implementation of this architecture by https://github.com/Janspiry/Palette-Image-to-Image-Diffusion-Models. Our U-Net uses three downsampling tiers, and each tier has two residual blocks. The first tier uses 128 inner channels, the second tier uses 256 inner channels, and the third tier uses 512 inner channels. The conditioning is passed into the network via channel concatenation. We train our network over the course of 6000 epochs. The learning rate scheduler follows cosine decay, starting with a base learning rate of 1e-4. The diffusion model scheduler uses 2000 denoising steps and a linear schedule that starts at 1e-6 and ends at 1e-2. The network is trained to minimize an $L_2$ loss. The network is trained using a dataset of 1533 samples and its performance is assessed during training using a holdout dataset of 192 samples.

#### A.2  Details of the inpainting networks

We train three conditional diffusion models to generate inflow planes from denoised lidar measurements. Each network handles inflow data for a different vertical extent. These diffusion models are also based on the Palette framework, though we modify the code to handle three-dimensional data via e.g., three-dimensional convolutions. The U-net architecture is the same architecture that is used for lidar denoising, except this network uses 96, 192, and 384 inner channels for its first, second, and third tiers, respectively. Due to memory limitations, the batch size is different for each of the diffusion models. The model that produces $z = [80, 120]$ m data has a batch size of 32, the model that produces $z = [0, 120]$ m data has a batch size of 16, and the model that produces $z = [0, 240]$ m data has a batch size of 8. The inflow networks are trained over 8000 epochs. They also use a cosine learning rate scheduler, starting with a base learning rate of 1e-4. The diffusion model scheduler uses 2000 denoising steps and a linear schedule that starts at 1e-6 and ends at 2.5e-2. The network is trained to minimize $L_2$ loss. The networks were trained using a dataset of 4350 samples, and their performance was assessed during training using a holdout dataset of 480 samples.